\documentclass[12pt]{article}
\usepackage[utf8]{inputenc}
\usepackage[russian,english]{babel} 
\usepackage[T2A]{fontenc}   
\usepackage{mathtools}
\usepackage{amsfonts,amssymb,amsmath} 
\usepackage{graphicx} 
\usepackage[noadjust]{cite}
\usepackage{hyperref}
\hypersetup{colorlinks=true, linkcolor=blue, filecolor=blue, urlcolor=blue}
\usepackage[table]{xcolor}
\definecolor{DefinedColor}{rgb}{0.9, 0.9, 0.95}

\newtheorem{statement}{Statement}

\newcommand{\om}{\omega}

\allowdisplaybreaks

\allowdisplaybreaks 
\usepackage{diagbox} 

\begin{document}

\title{Gradient Projection Method for Constrained Quantum Control}

\author{ Oleg~V.~Morzhin$^{1,2,}$\footnote{E-mail: \url{morzhin.oleg@yandex.ru};~ 
    \href{http://www.mathnet.ru/eng/person30382}{mathnet.ru/eng/person30382};~ 
    \href{https://orcid.org/0000-0002-9890-1303}{ORCID 0000-0002-9890-1303}} 
    \quad and \quad 
Alexander~N.~Pechen$^{1,2,}$\footnote{E-mail: \url{apechen@gmail.com};~ 
    \href{http://www.mathnet.ru/eng/person17991}{mathnet.ru/eng/person17991};~ 
    \href{https://orcid.org/0000-0001-8290-8300}{ORCID 0000-0001-8290-8300}} 
    \vspace{0.2cm} \\
\footnotesize $^1$ Department of Mathematical Methods for Quantum Technologies,, \vspace{-0.2cm} \\ 
\footnotesize Steklov Mathematical Institute of Russian Academy of Sciences, \vspace{-0.2cm} \\
\footnotesize Gubkina Str.~8, Moscow, 119991, Russia; \\
\footnotesize $^2$ University of Science and Technology MISIS, \vspace{-0.2cm}\\
\footnotesize Leninskiy Prospekt~4, Moscow, 119049, Russia}
\normalsize 

\date{}
\maketitle
\vspace{-0.8cm}

\begin{abstract}
In this work, we adopt the Gradient Projection Method (GPM) to problems of quantum control.\footnote{This paper develops GPM to various classes of dynamic quantum constrained optimization problems. The improved and published journal version is [{\it J.~Phys.~A: Math. Theor.} {\bf 58}, 135303 (2025)]. We  emphasize that a foundation of the first-order GPM for quantum constrained optimization was developed in the pioneering work [Oza~A., Pechen~A., Dominy~J., Beltrani~V., Moore~K., Rabitz~H. Optimization search effort over the control landscapes for open quantum systems with Kraus-map evolution. {\it J.~Phys.~A: Math. Theor.} {\bf 42}:20, 205305 (2009). \url{https://doi.org/10.1088/1751-8113/42/20/205305}] (culminating in Section~5 and Appendix~B) by considering finite-dimensional Kraus maps as explicit controls and parameterizing them by points of a suitable complex Stiefel manifold. The constrained optimization of quantum systems was formulated as optimization over complex Stiefel manifolds with additional constraints, where Kraus maps are used as explicit controls for manipulating quantum systems. Then, using this parametrization, the GPM was formulated for general quantum control systems with explicit derivations of the gradient of the objective functional and of the projection followed by the numerical analysis for some examples.}  For general $N$-level closed and open quantum systems, we derive the corresponding adjoint systems and gradients of the objective functionals, and provide the projection versions of the Pontryagin maximum principle and the GPM, all directly in terms of quantum objects such as evolution operator, Hamiltonians, density matrices, etc. Various forms of the GPM, including one- and two-step, are provided and compared. We formulate the GPM both for closed and open quantum systems, latter for the general case with simultaneous coherent and incoherent controls. The GPM is designed to perform local gradient based optimization in the case when bounds are imposed on the controls. The main advantage of the method is that it allows to exactly satisfy the bounds, in difference to other approaches such as adding constraints as weight to objective. We apply the GPM to several examples including generation of one- and two-qubit gates and two-qubit Bell and Werner states for models of superconducting qubits  under the constraint when controls are zero at the initial and final times, and steering an open quantum system state to a~target density matrix for simulating action of the Werner-Holevo channel, etc.

\vspace{0.1cm}


\end{abstract}

\section{Introduction} 

Quantum control is an important tool for development of various modern quantum technologies. Theory of optimal control of various quantum systems such as atoms, molecules, etc. forms an~important research direction at the intersection of mathematics, physics, chemistry, and computer science. It exploits diverse mathematics tools~\cite{KochEtAlEPJQuantumTechnol2022,KuprovIBook2023,DAlessandroBook2021,KurizkiKofmanBook2021,TannorBook2007,BrifNewJPhys2010,ButkovskiySamoilenkoBook,Gough2012,BoscainPRXQuantum2021} and has applications to controlled generation of quantum gates for closed~\cite{PalaoKosloff2002,GornovDvurechenskiiZarodnyukEtAl2011} and  open quantum systems~\cite{GoerzReichKochNJP2014,PechenPetruhanovMorzhinVolkov2024}, controlled transfer along a~spin chain~\cite{MurphyMontangeroGiovannettiCalarco2010}, nuclear magnetic resonance~\cite{KhanejaEtAlGRAPE2005,MaximovJChPh2008,KuprovIBook2023}, etc. Quantum systems in applications are typically open, i.e. interacting with their environment (reservoir). Control of closed quantum systems, which do not interact with an~environment (except possibly with {\it coherent control}), are also of interest. Incoherent  control approach for {\it $N$-level} open quantum systems with coherent control entering in the Hamiltonian and {\it incoherent control} entering  in the superoperator of dissipation was developed~\cite{PechenRabitzPRA2006,PechenPRA2011} and applied to various  quantum control problems for $N=2,3,4$ level quantum systems~\cite{PetruhanovPechenJPA2023,LokutsievskiyPechenZelikinJPA2024},~etc. 

Various optimization methods are used in optimal open-loop quantum control. For {\it infinite-dimensional}  optimization with various classes of functional controls one can apply, for example, the steepest-descent method (SDM)~\cite{GrossNeuhauserRabitz1992}, Pontryagin maximum principle (PMP)~\cite{ButkovskiySamoilenkoBook,GaronPRA2013,BoscainPRXQuantum2021,HansonLucarelliArXiv2024}, Krotov method (with or without regularization, etc.; \cite{TannorKazakovOrlov1992}, \cite[\S~16.2.2]{TannorBook2007}, \cite[p.~253--259]{KrotovBook1996}, \cite{SchmidtNegrettiAnkerholdCalarcoStockburger2011,GoerzReichKochNJP2014,MorzhinPechenProcSteklov2024}), Zhu--Rabitz method~\cite{ZhuRabitz1998} and the Turinici version~\cite{TuriniciIFAC2003} for bounded controls, 
Maday--Turinici~\cite{MadayTurinici2003,MaximovJChPh2008} and Ho--Rabitz~\cite{HoRabitzPRE2010} methods, 
GPM (one- and two- and three-step forms)~\cite{MorzhinPechenLJM2019,AbdullaRThesisMS2019,MorzhinPechenQIP2023,MorzhinPechenProcSteklov2024,MorzhinPechenEntropy2024}, conditional gradient method (CGM)~\cite{MorzhinPechenLJM2019}, speed gradient method~\cite{PechenBorisenokFradkov2022}. For {\it finite-dimensional}  optimization with various classes of parameterized controls one can use, e.g., 
GRadient Ascent Pulse Engineering (GRAPE) method (either for coherent control~\cite{KhanejaEtAlGRAPE2005,KuprovIBook2023} or for simultaneous coherent and incoherent controls~\cite{PetruhanovPechenJPA2023}), 
Broyden--Fletcher--Goldfarb--Shanno (BFGS) quasi-Newton 
GRAPE~\cite{deFouquieresSchirmerGlaserKuprov2011}, Chopped RAndom Basis (CRAB) ansatz~\cite{CanevaPRA2011,DoriaCalarcoMontangero2011}, genetic algorithm~\cite{JudsonRabitz1992,PechenRabitzPRA2006}, and machine learning in the form of AlphaZero deep exploration~\cite{DalgaardMotzoiSorensenSherson2020}. Combined optimization schemes are also used, e.g., \cite{GoerzWhaleyKoch2015} firstly using the Nelder--Mead (N.-M.) method for a~class of parameterized controls and the Krotov method starting after that. As \cite[\S~3.7, \S~6.8]{DykhtaSamsonyukBook2003}, 
\cite{GurmanRasinaBaturinaIFAC2013}, etc. show, such the branch of the mathematical theory of optimal control that is devoted to degenerate control problems and impulse control problems with discontinuous trajectories (\cite{GurmanBook1977, DykhtaSamsonyukBook2003}, etc.) can be used  for such a~quantum optimization problem that linearly contains unbounded functional control, and there may be one control with unbounded magnitudes and simultaneously another control with bounded magnitudes. A~deep neural network with the REINFORCE policy gradient algorithm was applied to control the state probability distribution of a~controllable closed quantum system~\cite{FouadST2024}. Several GPM versions were used in~\cite{BolducKneeGaugerLeachNature2017} for optimization over density matrices in quantum state tomography.

Often in quantum control some constraints, such as for example on absolute magnitudes of the controls~\cite{TuriniciIFAC2003,BoscainPRXQuantum2021,MorzhinPechenEntropy2024}, piecewise constant variations~\cite[p.~1537]{MorzhinPechenLJM2019}, spectrum~\cite{ReichPalaoKochJModOpt2014}, energy of the applied electromagnetic control pulse, should be taken into account for a quantum control problem. Handling such constraints may be done by  adding a weighted integral term to the objective functional. For example, if one aims to minimize the energy of the control pulse  then adding an integral term to the control objective can be an appropriate way. However, if one should impose some lower and upper bounds on the magnitudes of the control at each point of the whole time range, then adding the corresponding integral term to the objective does not guarantee exact satisfaction of these bounds. Setting some initial- and end-point constraints on the control (e.g., for continuously switching on coherent control from zero initial value and switching it off to the final zero value, e.g., in~\cite[p.~1897]{SundermanndeVivieRiedle1999}, \cite[p.~6]{HansonLucarelliArXiv2024}) is an another example when adding an integral term is not the optimal way. In this article, to take into account the initial-point, pointwise, and end-point constraints on magnitudes of the controls, we develop to such quantum control problems the GPM approach based on the fundamental theoretical works on one-step GPM (\cite{LevitinPolyak1966,DemyanovRubinovBook1970,FedorenkoBook1978,NikolskiiMS2007}, etc.), two-step GPM~\cite{AntipinDifferEqu1994} based on the heavy-ball method~\cite{PolyakUSSRComputMathMathPhys1964}, and three-step GPM~\cite{NedichIzvVUZov1993}. In quantum control problems, moreover, various state constraints represented via corresponding integral terms in objective functionals are considered~\cite{PalaoKosloffKochPRA2008,MorzhinPechenEntropy2024}. 

Going beyond the recent articles~\cite{MorzhinPechenQIP2023,MorzhinPechenEntropy2024,MorzhinPechenProcSteklov2024}, we adopt the GPM approach for optimal control of {\it $N$-level} quantum systems, with unitary and dissipative dynamics, with various objective functionals, to handle exactly various constraints. We consider control problems for general $N$-level closed quantum systems with coherent control, and for $N$-level open quantum systems, in full generality with both coherent control in the Hamiltonian and incoherent control in the dissipator (in some approaches, coherent control can appear in the dissipator as well~\cite{Aroch2024}). We derive the corresponding adjoint systems and gradients of the objective functional, using which provide the projection versions of the PMP and the GPM. All constructions are derived directly in terms of quantum objects such as evolution operator, Hamiltonians, density matrices, etc. 
Various forms of the GPM, including one- and two-step, are adopted and compared. The GPM is designed to perform local gradient based optimization in the case when bounds are imposed on the controls. The main advantage of the method is that it allows to exactly satisfy the constraints, in difference to other approaches such as adding constraints as a weight to the objective. GPM has the following properties: (1)~using the gradient of an~objective functional or function (also in SDM, CGM); (2)~operating in general with functional controls; (3)~allowing for an exact satisfaction with lower and upper pointwise bounds on magnitudes of controls (also in PMP, Krotov method, CGM), in contrast to possible combining, e.g., the exterior penalty method and SDM. To illustrate the operation of the method, we apply it to several examples including generation of one- and two-qubit gates and two-qubit Bell and Werner states for models of superconducting qubits~\cite{RigettiPRL2005} adding a constraint when controls are zero at the initial and final times, steering an open quantum system state to a target density matrix, e.g., for simulating action of the Werner-Holevo channel, etc. The simulations show that GPM-2 is able to {\it significantly accelerate} GPM-1 for the same~$\alpha$, and this effect is obtained only owing to the inertial term in GPM-2. In the optimization theory, an important issue is how to adapt the parameters of a used optimization method. In this  article, one of the developed GPM algorithms takes into account~\cite{MorzhinBuldaevVestnikBGU2008} and combines one-dimensional search for the step-size parameter at a~number of the one-step GPM first iterations and, after that, tries to adapt the parameter.

The  structure of the article is as follows. Sections~\ref{section2} and~\ref{section3} contain formulations of the GPM for solving constrained control problems for general $N$-level closed and open quantum systems, respectively, including description of the dynamical equations, objective functionals, constraints, derivation of the adjoint systems, gradients, PMP, and description of the GPM iterative processes. Section~\ref{section4} describes the GPM numerical results for the various control problems for $N=2,3,4$-level quantum systems, with comparing the  various GPM forms. Conclusions section~\ref{sectionConclusions} resumes the article. 

\section{Gradient projection method for closed quantum systems with coherent control}
\label{section2}

\subsection{Constrained control problems for closed quantum systems}
\label{subsection2.1}

A~closed $N$-level controlled quantum system evolves according to the Schr\"{o}dinger equation for the unitary evolution operator $U_t^u$: 
\begin{align}
\frac{dU_t^u}{dt} = -i \big( H_0 + \langle V(t), u(t) \rangle_{E^{N_u}} \big) U_t^u, \quad  U_{t=0}^u = \mathbb{I}_N,
\label{InitProblemForSchrodingerEq}
\end{align}
where coherent control $u$ with $N_u$ components is real-valued and considered here as piecewise continuous, $H_0$ is the free Hamiltonian, $V(t)$ is a family of Hamiltonians (continuous in $t$ fixed functions) describing interaction with control~$u$. We set Planck's constant $\hbar=1$. We consider the situation when the control~$u$ should satisfy the constraint
\begin{eqnarray}
u(t) \in Q_u(t) \subset \mathbb{R}^{N_u}, \quad \forall~t \in [0,T],
\label{GeneralPoinwiseConstraintForu}
\end{eqnarray}
where $Q_u(t)$ for each~$t$  is a~compact convex set. 

Consider minimization of some objective functional of the Mayer--Bolza type:
\begin{equation}
\Upsilon(u) = F(U_T^u) + \int\nolimits_0^T f(t, u(t))dt \to \inf, 
\label{UpsilonInf}
\end{equation} 
with some differentiable kinematic objective function~$F(U)$ and a differentiable in~$u$ function~$f(t,u)$. Particular physical examples include transfer to a~given state $\psi_{\rm target} \in \mathbb{CS}^{N-1}(1) \subset \mathbb{C}^N$ (unit sphere) and generating a~given unitary operation~$W_N$ under some control constraints. 

\begin{figure}[ht!]
\centering
\includegraphics[width=\linewidth]{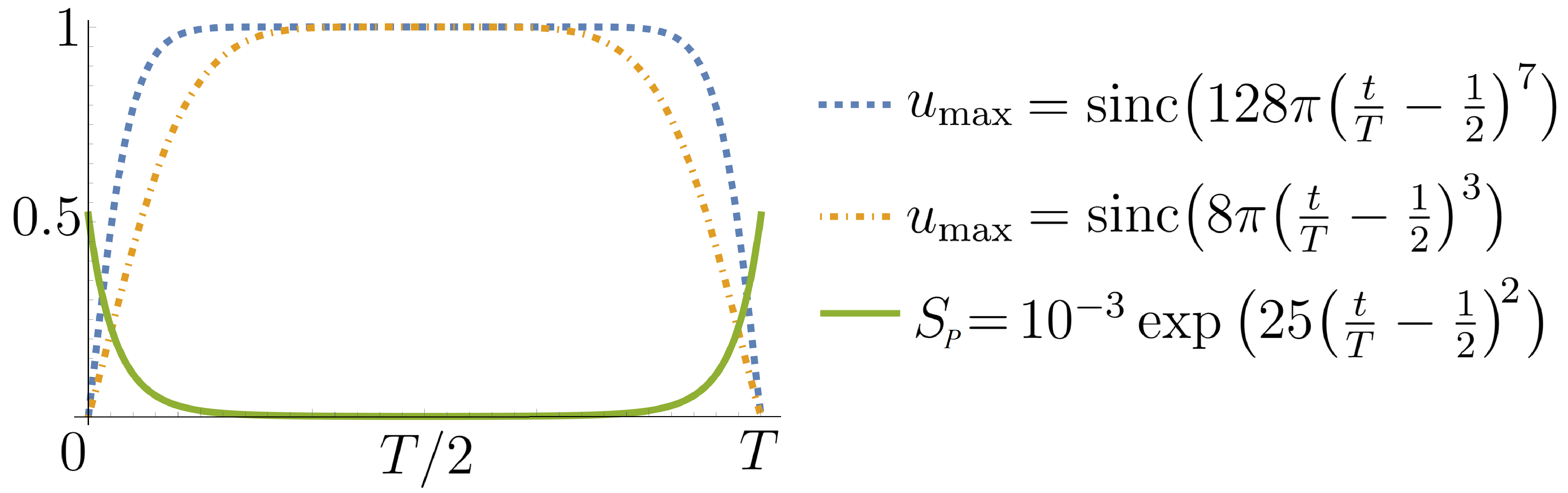}
\caption{\footnotesize For $C_{\max}^u = 1$, the two cases of the function $u_{\max}$,  correspondingly, with $q = 3, 7$. The function~$S_P = P \cdot S$ with $P = 10^{-3}$, $C_S = 25$.}
\label{MorzhinPechenFig1} 
\end{figure}

Examples of~(\ref{GeneralPoinwiseConstraintForu}) and for the initial-point, pointwise, and end-point constraints on magnitudes of controls, are given by
\begin{align}
u(t) \in Q_u(t) &:= [-u_{\max}(t), u_{\max}(t)]^{N_u} \quad {\rm s.t.}~
u_{\max}(t) = C_{\max}^u \,{\rm sinc}\Big(2^q \pi \Big(\frac{t}{T} - \frac{1}{2} \Big)^q \Big), 
\label{PointwiseConstraintForControl} \\
&\qquad q \in \{3,7\}, \quad {\rm sinc}(y) = 
\begin{cases}
\frac{\sin y}{y}, \quad y \neq 0, \\ 
1, \quad\quad\,  y = 0,
\end{cases} \nonumber
\end{align}
with the corresponding objective functionals 
\begin{equation}
\Upsilon_i(u) = J_i(u) + P \int\nolimits_0^T S(t) \| u(t) \|_{E^{N_u}}^2 dt \to \inf, \quad i = 1,2,
\label{UpsiloniInf}
\end{equation} 
where the terminal functionals for state transfer $|\psi_0\rangle\to |\psi_{\rm target}\rangle$ and gate $W$ generation, respectively, are 
\begin{align}
J_1(u) &=  F_1(U_T^u) = 1- \left| \langle \psi_{\rm target}, U_T^u \, \psi_0 \rangle \right|^2, 
\label{J1} \\
J_2(u) &=  F_2(U_T^u) = 1- \frac{1}{N^2} \left| {\rm Tr} (U_T^u \, W^{\dagger}) \right|^2.
\label{J2}
\end{align}
Here the boundary function $u_{\max}$ satisfies the condition $u_{\max}(0) = u_{\max}(T) = 0$ and is bounded above by a~given value $C_{\max}^u > 0$ so that $\max\limits_{t \in [0,T]} u_{\max}(t) = C_{\max}^u u_{\max}(T/2) = C_{\max}^u$. In applications, one can consider the penalty parameter $P \geq 0$ and the shape function~$S = \exp\Big(C_S \Big(\frac{t}{T} - \frac{1}{2}\Big)^2\Big)$ with $C_S > 0$. The integral term represents the requirement for the control~$u$ to continuously switch-on at $t=0$ and switch-off at $t=T$, in addition to be bounded by $u_{\max}$.  Each functional $\Upsilon_i$ represents the linear convolution of the criteria to minimize the functional $J_i$ and the penalty integral part. The functions $S_P = P \cdot S, ~u_{\max}$ are plotted in~Fig.~\ref{MorzhinPechenFig1} for some $C_{\max}^u, q, P, C_S$. 

\subsection{Adjoint systems, gradients, Pontryagin maximum principle, and GPM iterative processes}
\label{subsection2.2}

In this section, we derive the construction of the GPM for an iterative numerical solving of the considered minimization problems with the constraint~(\ref{GeneralPoinwiseConstraintForu}). For this, we use the Krotov Lagrangian~\cite{KrotovBook1996} at an admissible control~$u$ in the following specific form: 
\begin{align*}
L(u) &=  G(U_T^u) - \int\nolimits_0^T R(t, B_t^u, U_t^u, u(t))dt, \\
G(U_T^u) &=  F(U_T^u) + {\rm Re}\langle B_T^u, U_T^u \rangle - {\rm Re}\langle B_0^u, U_0^u \rangle, \\ 
R(t, B_t^u, U_t^u, u(t)) &=  {\rm Re}\Big[\big\langle B_t^u, -i \big( H_0 + \langle V(t), u(t) \rangle_{E^{N_u}} \big) U_t^u \big\rangle + \Big\langle \frac{d B_t^u}{dt}, U_t^u \Big\rangle \Big] - f(t, u(t)),
\end{align*}
where $B_t^u$ is a~unitary operator to be specified below. We also need the Pontryagin function $h(t, B, U, u) = {\rm Re} \big \langle B, -i \big( H_0 + \langle V(t), u \rangle_{E^{N_u}} \big) U \big\rangle - f(t,u)$ where $U,B$ are $N\times N$ unitary matrices and $u \in \mathbb{R}^{N_u}$. Consider the difference of $L$ at an arbitrary admissible~$u$ and a~given admissible~$u_k$: $L(u) - L(u_k) = G(U_T^u) - G(U_T^{u_k}) - \int\nolimits_0^T \big[ R(t, B_t^{u_k}, U_t^u, u(t)) - R(t, B_t^{u_k}, U_t^{u_k}, u_k(t)) \big] dt$ where $U^{u_k}_t$ is obtained by solving (\ref{InitProblemForSchrodingerEq}) when $u = u_k$ and $B_t^{u_k}$ is defined below. Using the Taylor matrix expansion for the increment of $G$ and equating the first order term to zero, we obtain the  transversality condition 
\begin{equation}
B_T^{u_k} = -\nabla F(U)|_{U = U_T^{u_k}}. 
\label{TransversalityConditionU}
\end{equation}
Making the Taylor expansion for the increment of $R$ and 
equating the first order term related to $U_t^u - U_t^{u_k}$ to zero, we obtain the dynamic equation 
\begin{equation}
\frac{d B_t^{u_k}}{dt} = -i \Big( H_0 + \langle V(t), u(t) \rangle_{E^{N_u}} \Big) B_t^{u_k}
\label{DifferentialEquationAdjointSystemB}
\end{equation} 
and the increment formula
$\Upsilon(u) - \Upsilon(u_k) = -\int\nolimits_0^T \sum\nolimits_{l=1}^{N_u} \big( \, {\rm Im} \big \langle B_t^{u_k},  V_l(t) U_t^{u_k} \big\rangle - \big(\frac{\partial f(t,u)}{\partial u}\big|_{u = u_k(t)}\big)_l \big) \big( u_l(t) - u_{k,l}(t) \big) dt + {\rm residual}$, in the class of admissible controls. The quantity $\delta_1(u_k) = \int\nolimits_0^T \big\langle {\rm grad}\, \Upsilon(u_k)(t), \, u(t) - u_k(t) \big\rangle dt$  characterises the increment of the objective functional. Using the increment formula, we obtain the gradient of $\Upsilon(u)$ at~$u_k$: 
\begin{align}
{\rm grad} \, \Upsilon(u_k)(t) &= -\frac{\partial h(t,B,U,u)}{\partial u}\Big|_{B=B_t^{u_k},~U=U_t^{u_k},~u=u_k(t)} \nonumber \\ 
&= \Big(- \, {\rm Im} \big\langle B_t^{u_k}, V_l(t) U_t^{u_k} \big\rangle + 
\big(\frac{\partial f(t,u)}{\partial u}\big|_{u = u_k(t)}\big)_l, ~ l = \overline{1,N_u} \Big).
\label{GradientofUpsilonukUB}
\end{align}

\begin{statement} (Gradient).
\label{statement1}
For the system~(\ref{InitProblemForSchrodingerEq}), the gradient of the objective functional~$\Upsilon(u)$ at an admissible control $u_k$ is given by~(\ref{GradientofUpsilonukUB}) and computed using the adjoint system~(\ref{TransversalityConditionU}),~(\ref{DifferentialEquationAdjointSystemB}).
\end{statement}

Consider the projection notation ${\rm Pr}_{[a,b]}(z)$ which means~$z$, if $z \in [a,b]$, means~$a$, if $z < a$, and means~$b$, if $z > b$. By analogy with~\cite[Eq.~(5)]{BuldaevKazminMathematics2022}, \cite[Proposition~3]{MorzhinPechenEntropy2024}, consider the following PMP form. 

\begin{statement} (PMP projection linearized version).
\label{statement2}
For the system~(\ref{InitProblemForSchrodingerEq}), control constraint~(\ref{GeneralPoinwiseConstraintForu}), and minimizing the objective in~(\ref{UpsilonInf}), if $u_k$ is a~local minimum point of $\Upsilon(u)$, then for $u_k$ there exist a solutions~$U_t^{u_k}$, $B_t^{u_k}$ such that the pointwise condition $u_k(t) = {\rm Pr}_{Q_u(t)}\big(u_k(t) - \alpha \, {\rm grad}\, \Upsilon(u_k)(t) \big)$ with $\alpha>0$ holds at $[0,T]$. 
\end{statement} 

In particular, for $\Upsilon_1(u)$, one has $B_T^{u_k} = \langle \psi_{\rm target}, U_T^{u_k} \psi_0 \rangle \psi_{\rm target} \psi_0^{\rm T}$ (matrix); for $\Upsilon_2(u)$, one has $B_T^{u_k} = \frac{1}{N^2} W_N {\rm Tr}(U_T^{u_k} W_N^{\dagger})$. For~(\ref{UpsiloniInf}), one has 
$\big(\frac{\partial f(t,u)}{\partial u}\big|_{u = u_k(t)}\big)_l = 2 P \, S(t) u_{k,l}(t)$ that is taken in the increment formula and in the gradient.  

Using the gradient~(\ref{GradientofUpsilonukUB}) and the constraint~(\ref{GeneralPoinwiseConstraintForu}), the following five GPM iterative processes starting from a~given initial guess~$u_0$ satisfying~(\ref{GeneralPoinwiseConstraintForu}) are formulated. 

(I) GPM-1($\alpha_k$, 1dCOM) is the one-step one-parameter GPM  such that
\begin{align*}
u_{k+1}(t) &= u_{k,\alpha_k}(t), \quad  \alpha_k = {\rm arg}\min\limits_{\alpha \in [\underline{\alpha}, \overline{\alpha}]} \Upsilon(u_{k,\alpha}), \quad k \geq 0, \\ 
u_{k,\alpha}(t) &= {\rm Pr}_{Q_u(t)}\big(u_k(t) + \alpha \, {\rm grad} \, \Upsilon(u_k)(t) \big), \quad \alpha > 0, \quad t \in [0, T],
\end{align*}
so that the step-size parameter $\alpha$ should be optimized (globally) at each iteration at $[\underline{\alpha}, \overline{\alpha}]$ with given $0 \leq \underline{\alpha} < \overline{\alpha}$. This GPM is given using \cite[\S~18]{FedorenkoBook1978} and Scheme~3 from \cite[\S~IV.2]{DemyanovRubinovBook1970}. ``1dCOM'' means an appropriate one-dimensional constrained optimization method.

(II) GPM-1(fixed~$\alpha$) is the one-step one-parameter GPM such that $u_{k+1}(t) = u_{k,\alpha}(t)$ with $\alpha$ fixed for the whole allowed number of iterations. This GPM is given using \cite[\S~IV.2, Scheme~2]{DemyanovRubinovBook1970}, \cite{BuldaevKazminMathematics2022}, \cite[\S~3.3]{BuldaevChapterSpringer2010}. 

(III) GPM-1($\alpha_k$, adapt., 1dCOM) is the one-step one-parameter GPM  such that  at its first iterations with $k = \overline{0,\widehat{k}-1}$ uses 1dCOM for obtaining $\alpha_k$ and, after that, works as follows. If $k=\widehat{k}$, then GPM takes the parameter $\alpha_k = \Big(\prod_{k=0}^{\widehat{k}-1} \alpha_k \Big)^{1/\widehat{k}}$ (geometric mean). If $k>\widehat{k}$ and $\frac{100 (\Upsilon(u_k) - \Upsilon^{\rm best})}{\Upsilon(u_k)} > y$ [\%] ($y$ is preliminary defined), then $\alpha_k$ is changed by the formula $\alpha_k := \theta \, \alpha_k + (1 - \theta) \tau \alpha_k$, where the convex combination's parameter $\theta \in (0,1)$ is predefined and the parameter $\tau = \frac{\Upsilon^{\rm best}}{\Upsilon(u_k)}$ (it assumes that $\Upsilon(u_k)>0$) and GPM takes the control $u^{\rm best}$ instead of the worse control~$u_k$. Here $\Upsilon^{\rm best}$ and $u^{\rm best}$ mean the current least value of $\Upsilon(u)$ and the corresponding control. This GPM is based on~\cite{MorzhinBuldaevVestnikBGU2008}.

(IV) GPM-2(fixed~$\alpha, \beta$) is the two-step two-parameter GPM  such that GPM-1(fixed $\alpha$) is used for $k=0$ and, after that, the following is used:
\[
u_{k,\alpha,\beta}(t) = {\rm Pr}_{Q_u(t)}\big(u_k(t) + \alpha \, {\rm grad} \, \Upsilon(u_k)(t) + \beta \left( u_k(t) - u_{k-1}(t) \right) \big), 
\]
where $\alpha >0$ and $\beta \in (0,1)$ are fixed for the whole allowed number of iterations. This GPM is constructed by analogy with the recent works~\cite{MorzhinPechenProcSteklov2024,MorzhinPechenEntropy2024} and is based on the heavy-ball method~\cite{PolyakUSSRComputMathMathPhys1964} and the projection version~\cite{AntipinDifferEqu1994}. 

(V) GPM-2($\alpha_k, \beta_k$, 2dCOM) is the two-step two-parameter GPM  such that GPM-1($\alpha_k$, 1dCOM) is used when $k=0$ and, after that, the following is used: 
\[
u_{k+1}(t) = u_{k,\alpha_k, \beta_k}(t), \quad  (\alpha_k, \beta_k) = {\rm arg}\min\limits_{(\alpha,\beta) \in [\underline{\alpha},\overline{\alpha}] \times [\underline{\beta}, \overline{\beta}]} \Upsilon(u_{k,\alpha,\beta}), \quad k \geq 0, 
\] 
where $0 \leq \underline{\beta} < \overline{\beta}<1$. This method is formulated by analogy with GPM-1($\alpha_k$, 1dCOM) and the given in~\cite[p.~60]{KhlebnikovBalashovTrembaBook2024} note to optimize $\alpha_k, \beta_k$ for modifying the heavy-ball method in the theory of finite-dimensional optimization.

In addition to a~given maximal allowed number of GPM iterations, as the other stopping conditions for GPM one can use  the conditions
\begin{eqnarray}
&J(u_k) < \varepsilon_{{\rm stop},J}~~\&~~\Upsilon(c_k) < \varepsilon_{{\rm stop}, \Upsilon}, 
\quad 0<\varepsilon_{{\rm stop},J}, \varepsilon_{{\rm stop}, \Upsilon} \ll 1, 
\label{StoppingCondition1} \\
&|\Upsilon(u_k) - \Upsilon(u_{k-1})| < \varepsilon_{{\rm stop}, \Delta \Upsilon} \ll 1.
\label{StoppingCondition2}
\end{eqnarray} 

For a qualitative comparison of these five GPM versions, we note the following. 

GPM-1($\alpha_k$, 1dCOM) and GPM-2($\alpha_k, \beta_k$, 2dCOM) represent the idea to look for the best possible decreasing for the objective functional at each iteration at the cost of optimizing $\alpha$ in the one-step GPM (by analogy with SDM) and $(\alpha, \beta)$ in the two-step GPM at each iteration in the sense of minimizing the corresponding objective functions $\Upsilon(u_{k,\alpha})$ and $\Upsilon(u_{k,\alpha,\beta})$. For GPM-1($\alpha_k$, 1dCOM), we note that for  a $u_k$ which does not satisfy PMP, if $\underline{\alpha}=0$ and 1dCOM, etc. works well, then $\Upsilon(u_{k+1}) < \Upsilon(u_k)$. For GPM-2($\alpha_k, \beta_k$, 2dCOM), such the property can also be given. The objective functions $\Upsilon(u_{k,\alpha})$, $\Upsilon(u_{k,\alpha,\beta})$ are determined implicitly through the dynamical system and objective functional and it may turn out that they are non-convex. Thus, the complexity of solving such a~one- or two-dimensional constrained optimization problem with high precision may turn out to be high. However, in such a parametric search, one can take such $\alpha$ or $(\alpha, \beta)$ that gives an~appropriate decrease of the objective functional at the current iteration of GPM, i.e. without requiring to optimize $\alpha$ or $(\alpha, \beta)$ with high precision. As 1dCOM, one may try, e.g., the generalized simulated annealing (GSA)~\cite{KirkpatrickGelattVecchi1983,TsallisStariolo1996,SciPyFunctionDualAnnealing} or N.-M. method~\cite{NelderMead1965,GaoHan2012,SciPyFunctionMinimizeNelderMead},~etc. Despite the possible difficulties in solving the problems of parametric optimization, these two GPM versions are interesting, especially when use of GPM-1(fixed $\alpha$) or GPM-2(fixed $\alpha, \beta$) encounters difficulties in setting $\alpha$ or~$(\alpha, \beta)$.

For GPM-1(fixed $\alpha$) and GPM-2(fixed $\alpha, \beta$), one may expect that the values $\{\Upsilon(u_k)\}$ sequentially computed by one of these GPM versions may be not always changed monotonically and that, as a~variant, such a GPM may stop moving towards a~solution. However, these GPM forms are of interest, because, their structure is simple and, as the computations described  in~\cite{MorzhinPechenEntropy2024,MorzhinPechenProcSteklov2024} and below show, they can be practically useful. For GPM-2(fixed $\alpha, \beta$), relying on the known computational facts for the heavy-ball method in finite-dimensional optimization, e.g., in~\cite{SutskeverMartensDahlHinton2013}, we consider the inertial parameter $\beta \in (0, 1)$ and first of all $\beta = 0.8, 0.9$ should be~tried. GPM-1($\alpha_k$, adapt., 1dCOM) combines GPM-1($\alpha_k$, 1dCOM) and GPM-1(fixed $\alpha$) in a specific way.

A strategy of using these various GPM forms is that one initially tries GPM-2(fixed $\alpha,\beta$) or GPM-1(fixed $\alpha$) with some adjusting of the parameters taken fixed for some allowed numbers of iterations, with various initial guesses, and, after that, decides whether it is needed to use an another GPM form, etc. 

If the requirement for coherent control to continuously switch-on at $t=0$ and switch-off at $t=T$ is taken in the sense of (\ref{PointwiseConstraintForControl}), then each $u_{k+1}$ given by anyone of the considered GPM versions satisfies the condition $u_{k+1}(0) = u_{k+1}(T) = 0$ (in the computer implementation, with some high enough precision). 

\section{Gradient projection method for open quantum systems with coherent and incoherent controls}
\label{section3}

\subsection{Constrained control problems for open quantum systems}
\label{subsection3.1}

In this section, following~\cite{PechenRabitzPRA2006,PechenPRA2011} consider an open $N$-level quantum system with Markovian dynamics depending on coherent control~$u$ and incoherent control~$n$ consisting of $N_n$ components: 
\begin{align}
\frac{d\rho_t^c}{dt} &= -i [H_c, \rho_t^c] + \varepsilon \mathcal{D}_n(\rho_t^c), \quad \rho_{t=0}^c = \rho_0, 
\label{OpenNlevelQuantumSystem} \\
H_c &= H_0 + \varepsilon H_n + \sum\nolimits_{l=1}^{N_u} H_l u_l, \quad \mathcal{D}_{n(t)}(\rho_t^c) = \sum\nolimits_{s=1}^{N_\gamma} \gamma_s(t) \, \mathcal{D}_s(\rho_t^c), 
\label{OpenSystemGeneralHamiltonianDissipator}
\end{align}
where $t \in [0,T]$ with a~given $T>0$, $\rho_t^c$ is the system density matrix, 
$H_n$ is the controlled effective Hamiltonian, 
$\{\gamma_s\}$ are generally time-dependent decoherence rates which may depend on incoherent control ($\gamma_s(t) = \gamma_s(n(t))$) and in this case which are assumed to be  differentiable in~$n$, $\mathcal{D}_n(\rho_t^c)$ is the controlled superoperator of dissipation, $\mathcal{D}_s(\rho_t^c)$ are some fixed superoperators, the initial density matrix~$\rho_0$ and coupling strength $\varepsilon>0$ are given, real-valued control $c=(u,n)$ is assumed to be piecewise continuous. Assume the control $c$ should satisfy the constraint 
\begin{equation}
c(t) = (u(t), n(t)) \in Q_c(t) \subset \mathbb{R}^{N_u + N_n}, \quad \forall~t \in [0,T],
\label{GeneralPoinwiseConstraintForc}
\end{equation}
where $Q_c(t)$ is a~compact convex set which also takes into account the physical bounds on the incoherent control~$n$,  $n_l(t) \geq 0$ for each $l$th component of~$n$. Consider the objective functional of the Mayer--Bolza type:
\begin{equation}
\Theta(c) = F(\rho_T^c) + \int\nolimits_0^T \big[ g(t,\rho_t^c) + f(t, c(t)) \big] dt \to \inf, 
\label{ThetaOpenInf}
\end{equation} 
where a~given differentiable kinematic objective function~$F(\rho)$ and the integral term with a~given differentiable in~$\rho$ function~$g(t,\rho)$ and  differentiable in~$u$ function~$f(t,u)$ are specified~below. 

We specify the  superoperator of dissipation as:
\begin{align}
\mathcal{D}_{n(t)}(\rho_t^c) &= \sum\nolimits_{j=1}^{N_n} \Big(\Omega_j (n_j(t)+1) (2 A_j \rho_t^c A_j^{\dagger} - \{ A_j^{\dagger} A_j, \rho_t^c \}) \nonumber \\
&\qquad + \Omega_j n_j(t) (2 A_j^{\dagger} \rho_t^c A_j - \{ A_j A_j^{\dagger}, \rho_t^c \})\Big), 
\label{CertainDissipator}
\end{align}
where $\Omega_j>0$ and the Lindblad operators $A_j$ are given. In terms of (\ref{OpenSystemGeneralHamiltonianDissipator}), we have $N_{\gamma} = 2 N_n$ and 
\begin{align*}
\gamma_s(t) &=  \begin{cases}
n_j(t)+1, & {\rm if~}s~{\rm is~odd}, \\
n_j(t), & {\rm if~}s~{\rm is~even}, \end{cases} \\
\mathcal{D}_s(\rho_t^c) &= \begin{cases} 
\Omega_j (2 A_j \rho_t^c A_j^{\dagger} - \{ A_j^{\dagger} A_j, \rho_t^c \}), & {\rm if~}s~{\rm is~odd}, \\
\Omega_j (2 A_j^{\dagger} \rho_t^c A_j - \{ A_j A_j^{\dagger}, \rho_t^c \}), & {\rm if~}s~{\rm is~even}. \end{cases}
\end{align*}  

An example is a constraint
\begin{equation}
c(t) = (u(t),n(t)) \in Q_c := [-C_{\max}^u, C_{\max}^u]^{N_u} \times [0, C^n_{\max}]^{N_n} \quad \forall t \in [0,T],
\label{ParallelepipedalConstraintControlcOpen}
\end{equation}
with some given $C_{\max}^u, C^n_{\max} > 0$. 

For formulating examples of (\ref{ThetaOpenInf}), we specify $F(\rho_T^c)$, $g(t,\rho_t^c)$, and $f(t,c(t))$. Consider 
\begin{equation}
f(t,c(t)) = P_u \, S(t) \,  \sum\nolimits_{l=1}^{N_u} u_l^2(t) + P_n \sum\nolimits_{l=1}^{N_n} n_l(t), 
\label{certainftc}
\end{equation}
where the shape function $S$ was specified above, the penalty parameters $P_u, P_n \geq 0$ are given. Below the target density matrix~$\rho_{\rm target}$, the penalty parameter $P_{\rho} >0$, weighting non-decreasing function~$\sigma$ of $t$ are given. The Hilbert--Schmidt scalar product (overlap) $\langle \rho_1, \rho_2 \rangle = {\rm Tr}(\rho_1 \rho_2)$ and squared Hilbert--Schmidt distance $\| \rho_1 - \rho_2 \|^2 = {\rm Tr}((\rho_1 - \rho_2)^2)$ for density matrices $\rho_1, \rho_2$ are used. Consider the following three control problems.
\begin{itemize}
\item Steering to $\rho_{\rm target}$ under the state constraint $d(t,\rho_t^c) \equiv 0$ taken into account as an  integral term. We specify $\Theta_1(c)$ taking $F(\rho_T^c) = \| \rho_T^c - \rho_{\rm target} \|^2$ and $g(t,\rho_t^c) = P_{\rho} \, d(t, \rho_t^c)$ and selecting either  $d(t,\rho_t^c) = \sigma(t) \| \rho_t^c - \rho_{\rm target} \|^2$ (weighted closeness function) or $d(t,\rho_t^c) = \| \rho_t^c - \overline{\rho}_t \|^2$ where $\overline{\rho}_t$ is some density matrix continuous function satisfying $\overline{\rho}_{t=0} = \rho_0$, $\overline{\rho}_{t=T} = \rho_{\rm target}$. With this $F(\rho_T^c)$, we define the terminal functional $I_1(c)$. 

\item Keeping the system at $\rho_0$ at the whole $[0,T]$, i.e. $\| \rho_t^c - \rho_0\| \equiv 0$. We specify $\Theta_2(c)$ taking $F(\rho_T^c) = \|\rho_T^c - \rho_0\|^2$ and $g(t,\rho_t^c) = P_{\rho} \, \| \rho_t^c - \rho_0 \|^2$. With this  $F(\rho_T^c)$, we define the terminal functional $I_2(c)$. 

\item Maximizing the final Hilbert--Schmidt overlap (scalar product) between $\rho_T^c$ and $\rho_{\rm target}$ under the corresponding weighted integral closeness term. We specify $\Theta_3(c)$ taking  $F(\rho_T^c) = b - \langle \rho_T^c, \rho_{\rm target} \rangle$ and $g(t,\rho_t^c) = P_{\rho} \, \sigma(t) \big(b - \langle \rho_t^c, \rho_{\rm target} \rangle \big)$ where the upper bound $b$ is given. With this  $F(\rho_T^c)$, we define the terminal functional~$I_3(c)$. 
\end{itemize}

Further, consider the problem of generating a unitary $p$-qubit $(N=2^p)$ gate~$W_N$ in an open system of the type~(\ref{OpenNlevelQuantumSystem}), (\ref{OpenSystemGeneralHamiltonianDissipator}). Following the Goerz--Reich--Koch approach~\cite{GoerzReichKochNJP2014} and using both coherent and incoherent controls, we consider
\begin{equation}
I_4(c) = \frac{1}{6}\sum\nolimits_{m = 1}^3 \| \rho^c_{T,m} - W_N \rho_{0,m} W_N^{\dagger} \|^2 \to \inf, 
\label{GRKI4inf}
\end{equation} 
where $\{\rho_{0,m}\}$ are the special initial matrices given in (4a)--(4c) in~\cite{GoerzReichKochNJP2014}: 
$(\rho_{1,0})_{ij} = \frac{2(N-i+1)}{N(N+1)}\delta_{ij}$, $(\rho_{2,0})_{ij} = \frac{1}{N}$, $(\rho_{3,0})_{ij} = \frac{1}{N}\delta_{ij}$ with the Kronecker delta $\delta_{ij}$ and $i,j = \overline{1,N}$; the final states $\rho^c_{T,m}$, $m=1,2,3$ correspond to these three initial states. 
In contrast to~(\ref{ThetaOpenInf}) and the specified above three one-state problems, here we operate with the triple of the density matrices of the system~(\ref{OpenNlevelQuantumSystem}) taken with the three special initial states. The article~\cite{PechenPetruhanovMorzhinVolkov2024} uses~(\ref{GRKI4inf}) for the case $N=4$ and consider the C-PHASE, CNOT, SWAP gates and piecewise constant coherent and incoherent controls and the adaptations of GRAPE and dual~annealing. Adding the term $P_u \int\nolimits_0^T S(t) \| u(t) \|^2_{E^{N_u}} dt$ to $I_4(c)$ defines~$\Theta_4(c)$. 

\subsection{Adjoint systems, gradients, Pontryagin maximum principle, and GPM iterative processes}
\label{subsection3.2}

Here we adapt GPM for iterative numerical solving of the one-state problem~(\ref{ThetaOpenInf}) with the constraint~(\ref{GeneralPoinwiseConstraintForc}). We use the Krotov Lagrangian~\cite{KrotovBook1996} at an admissible control~$c$ in the following specific form: 
\begin{align*}
L(c) &= G(\rho_T^c) - \int\nolimits_0^T R(t, \chi_t^c, \rho_t^c, c(t))dt, \\
G(\rho_T^c) &= F(\rho_T^c) + \langle \chi_T^c, \rho_T^c \rangle - \langle \chi_0^c, \rho_0 \rangle, \\
R(t, \chi_t^c, \rho_t^c, c(t)) &= \langle \chi_t^c, -i [H_{c(t)}, \rho_t^c] + \varepsilon \mathcal{D}_{n(t)}(\rho_t^c) \rangle + \Big\langle \frac{d\chi_t^c}{dt}, \rho_t^c \Big\rangle - g(t,\rho_t^c) - f(t,c(t)),
\end{align*}
where $\chi_t^c$ is a density matrix to be specified. We will also use the Pontryagin function $h(t, \chi, \rho, c) = \big\langle \chi, -i [H_c, \rho] + \varepsilon \mathcal{D}_n(\rho) \big\rangle - g(t,\rho) - f(t,c)$ where $\rho,\chi$ are $N\times N$ density matrices and $c \in \mathbb{R}^{N_u + N_n}$. Operating with the increment $L(c) - L(c_k)$ by analogy with \cite{MorzhinPechenEntropy2024} and subsection~\ref{subsection2.2} gives the following adjoint system and gradient at $c_k$:
\begin{align}
&\frac{d\chi_t^{c_k}}{dt} = -i [H_{c_k}, \chi_t^{c_k}] - \varepsilon \mathcal{D}_{n_k}^{\dagger}(\chi_t^{c_k}) + 
\frac{\partial g(t,\rho)}{\partial \rho}\Big|_{\rho = \rho_t^{c_k}}, \quad 
\chi_T^{c_k} = -\frac{d F(\rho)}{d\rho}\Big|_{\rho = \rho_T^{c_k}}, 
\label{AdjointSystemOpen} \\
&{\rm grad} \, \Theta(c_k)(t) = \Big(\left\langle \chi_t^{c_k}, \, i[H_l,\rho^{(k)}(t)] \right\rangle + 
\frac{\partial f(t,c)}{\partial u_l}\Big|_{c=c_k(t)}, ~~ l = \overline{1,N_u}, \nonumber \\
&\qquad \Big\langle \chi_t^{c_k}, i \varepsilon \Big[\frac{\partial H_n}{\partial n_l}\Big|_{n = n_k(t)}, \rho_t^{c_k} \Big] \Big\rangle + \varepsilon \frac{\partial \mathcal{D}_n(\rho)}{\partial n_l}\Big|_{n = n_k(t)} + \frac{\partial f(t,c)}{\partial n_l}\Big|_{c=c_k(t)}
,~ l = \overline{1,N_n} \Big).
\label{GradientofThetackRhoChi}
\end{align}

\begin{statement} (Gradient of $\Theta$ at $c_k$).
\label{statement3} 
For the system~(\ref{OpenNlevelQuantumSystem}) with (\ref{OpenSystemGeneralHamiltonianDissipator}), gradient of the objective functional $\Theta(c)$ at an admissible control $c_k$ is given by~(\ref{GradientofThetackRhoChi}) and is computed using the adjoint system~(\ref{AdjointSystemOpen}).  
\end{statement} 

\begin{statement} (PMP projection linearized version for the problem with the one-state objective functional $\Theta(c)$).
\label{statement4} 
For the system~(\ref{OpenNlevelQuantumSystem}) with (\ref{OpenSystemGeneralHamiltonianDissipator}), control constraint~(\ref{GeneralPoinwiseConstraintForc}), and the control goal~(\ref{ThetaOpenInf}), if $c_k$ is a~local minimum point of $\Theta(c)$, then for $c_k$ there exist the solutions~$\rho_t^{c_k}$, $\chi_t^{c_k}$ such  that the pointwise condition $c_k(t) = {\rm Pr}_{Q_c(t)}\big(c_k(t) - \alpha \, {\rm grad}\, \Theta(c_k)(t) \big)$ with $\alpha>0$ holds at $[0,T]$.  
\end{statement}

For the applied to $\rho_t^c$  superoperator of dissipation in the form~(\ref{CertainDissipator}), the applied to $\chi_t^{c_k}$ superoperator has the form
\begin{align*}
\mathcal{D}_{n_k(t)}^{\dagger}(\chi_t^{c_k}) &= \sum\nolimits_{j=1}^{N_n} \Big(\Omega_j (n_{k,j}(t)+1) (2 A_j^{\dagger} \chi_t^{c_k} A_j - \{ A_j^{\dagger} A_j, \chi_t^{c_k} \}) \\
&\qquad + \Omega_j n_{k,j}(t) (2 A_j \chi_t^{c_k} A_j^{\dagger} - \{ A_j A_j^{\dagger}, \chi_t^{c_k} \}) \Big), 
\end{align*}
where some part of the positions of $A_j, A_j^{\dagger}$ differs from the positions of $A_j, A_j^{\dagger}$ in $\mathcal{D}_n(\rho_t^c)$. In particular, for $F(\rho) = \| \rho - \rho_{\rm target} \|^2$, one has $\chi_T^{c_k} = -2(\rho_T^{c_k} - \rho_{\rm target})$; for $F(\rho) = b - \langle \rho, \rho_{\rm target} \rangle$, one has $\chi_T^{c_k} = \rho_{\rm target}$. For $g(t,\rho) = P_{\rho} \sigma(t) \| \rho - \rho_{\rm target} \|^2$, one has $\frac{\partial g(t,\rho)}{\partial \rho}\Big|_{\rho = \rho_t^{c_k}} = 2 P_{\rho} \sigma(t) (\rho_t^{c_k} - \rho_{\rm target})$; 
for $g(t,\rho) = P_{\rho} \| \rho - \overline{\rho}_t \|^2$, one has $\frac{\partial g(t,\rho)}{\partial \rho}\Big|_{\rho = \rho_t^{c_k}} = 2 P_{\rho} (\rho_t^{c_k} - \overline{\rho}_t)$; for $g(t,\rho) = P_{\rho} \sigma(t) \big( b - \langle \rho, \rho_{\rm target} \rangle \big)$, one has $\frac{\partial g(t,\rho)}{\partial \rho}\Big|_{\rho = \rho_t^{c_k}} = - P_{\rho} \sigma(t) \rho_{\rm target}$. For $f(t,c)$ defined in~(\ref{certainftc}), one has $\frac{\partial f(t,c)}{\partial u_l}\big|_{c=c_k(t)} = 2 P_u \, S(t) \, u_{k,l}(t)$ and $\frac{\partial f(t,c)}{\partial n_l}\big|_{c=c_k(t)} = P_n$. 

The Krotov Langangian, adjoint system, and gradient for the three-state problem with the objective functional $\Theta_4(c)$ are derived easily by analogy with the provided  above results for the one-state problem with $\Theta(c)$ and are omitted here for brevity.

Any GPM iterative process described above for the non-dissipative dynamics can be used here after the obvious substitutions, i.e. taking~$\Theta(c_k)$ instead of $\Upsilon(u_k)$, etc., including for $\Theta_4(c)$. Moreover, by analogy with \cite{MorzhinPechenProcSteklov2024} in quantum control and with  \cite{NedichIzvVUZov1993} in the theory of finite-dimensional optimization, consider GPM-3(fixed $\alpha, \beta, \xi$) where GPM-1(fixed $\alpha$) is used for $k=0$ and GPM-2(fixed $\alpha, \beta$) is used for $k=1$ and, after that, the following is used:
\begin{align*}
c_{k+1}(t) &= {\rm Pr}_{Q_c(t)}\big(c_k(t) + \alpha \, {\rm grad} \, \Theta(c_k)(t) + \beta \big( c_k(t) - c_{k-1}(t) \big) \nonumber \\
&\qquad + \xi \big( c_{k-1}(t) - c_{k-2}(t) \big)\big), \quad \alpha>0, ~\beta \in (0,1),~ 0<\xi<\beta.
\end{align*}

For stopping the iterations, one should take into account the computed values not only of~$\Theta$, but also of its terminal and state integral terms.

\section{Numerical simulations}
\label{section4}

\subsection{Real-valued versions and numerical implementation}
\label{subsection4.1}

Both for closed and open systems, one may use GPM either with the given above adjoint systems, gradients or for the corresponding optimal control problems with real-valued states. In this article, under piecewise linear interpolation for controls at a~uniform grid at $[0,T]$, we use the {\tt SciPy} function {\tt solve\_ivp}~\cite{SciPySolveIVP} for solving the Cauchy problems with real-valued states both for GPM designed without deriving the optimal control problems with real-valued states and for GPM designed after this deriving. Each given above iterative GPM formula contains ${\rm grad}\, \Upsilon(u_k)(t)$ or ${\rm grad}\, \Theta(c_k)(t)$ and is general both in the original terms ($U_t^u, V_t^u$ or $\rho_t^c, \chi_t^c$) and in the real-valued terms. Here we omit the adjoint systems, gradients for the problems with real-valued states both for brevity and due to the presence of the appropriate general adjoint system and gradient formulas in the real-valued terms in the theory of optimal control, e.g., the gradient~(10), adjoint system (12),~(13) in~\cite[\S~6.3]{VasilievFPBook1974}, the gradient formula~(2.5.29) in~\cite{PolakBook1971}. As an example, consider the general form of the bilinear system with real-valued states for~(\ref{OpenNlevelQuantumSystem}) as $\qquad\qquad\qquad\qquad\qquad\qquad \frac{dx_t}{dt} = \Big(A + \sum\nolimits_{l=1}^{N_u} B_{u,l} u_l(t) + \sum\nolimits_{l=1}^{N_n} B_{n,l} n_l(t) \Big) x_t$ with $x_{t=0}=x_{\rho_0}$. 

Below in GPM with the problems of optimizing $\alpha$ or $(\alpha,\beta)$, the corresponding objective functions are given implicitly and one should expect that they can be non-convex. Our numerical experiments use GPM-1($\alpha_k$, 1dCOM), etc. with N.-M. method or GSA as 1dCOM, 2dCOM with the {\tt SciPy} function {\tt minimize} setting {\tt method='Nelder-Mead'}~\cite{SciPyFunctionMinimizeNelderMead} and GSA as the {\tt SciPy} function {\tt dual\_annealing}~\cite{SciPyFunctionDualAnnealing} with the turned off optional local~search. 

\subsection{Generating the Hadamard gate for a~one-qubit supercondiucting system}
\label{subsection4.2}

Consider a dynamical model which represents a superconducting qubit manipulated by a~quasiperiodic driving~\cite{RigettiPRL2005}. Let ${\rm X} = 
\begin{pmatrix} 0 & 1 \\ 1 & 0 \end{pmatrix}$, ${\rm Z} = \begin{pmatrix} 1 & 0 \\ 0 & -1 \end{pmatrix}$, ${\rm Y} =  
\begin{pmatrix} 0 & -i \\ i & 0 \end{pmatrix}$ be Pauli matrices. In~(\ref{InitProblemForSchrodingerEq}), take  control $u = (u_x, u_y)$ and 
\[
H_0 = \Omega \, {\rm Z}, \quad \langle V(t), u(t) \rangle_{E^2} = 2(u_x(t) \cos(\omega t) + u_y(t)\sin(\omega t)) {\rm X}, 
\]
i.e. $V_1(t) = 2 \cos(\omega t) {\rm X}$, $V_2(t) = 2 \sin(\omega t) {\rm X}$. For this dynamical model, consider objective $J_2(u)$ for generating the Hadamard gate $W^H = \frac{1}{\sqrt{2}} \begin{pmatrix} 1 & 1 \\ 1 & -1 \end{pmatrix}$. Take $T=1.5$, $\Omega = \omega = 1$, and consider the following five~cases: 
\begin{itemize}
\item Case 1: $u_0=0$, $P=0$, without~(\ref{PointwiseConstraintForControl}); 

\vspace{-0.15cm}

\item Case 2: $u_0=0$, $P=0$, but (\ref{PointwiseConstraintForControl}) is used where $C_{\max}^u = 1$; 

\vspace{-0.15cm}

\item Case 3: $u_0=0$, $P = 10^{-3}$, $C_{\max}^u=1$; 

\vspace{-0.15cm}

\item Case 4: $u_0=0$, $P = 8 \cdot 10^{-3}$, $C_{\max}^u = 0.6$; 

\vspace{-0.15cm}

\item Case 5: $u_0 = -{\rm sinc}\Big(8 \pi \Big(\frac{t}{T} - \frac{1}{2} \Big)^3 \Big)$, $P=0$, without~(\ref{PointwiseConstraintForControl}). 
\end{itemize}

For these cases, below we compare various GPM forms. We take $\varepsilon_{{\rm stop},J} = 10^{-5}$, $\varepsilon_{{\rm stop}, \Upsilon} = 10^{-3}$, 
and $\varepsilon_{{\rm stop}, \Delta \Upsilon} = 10^{-8}$ in~(\ref{StoppingCondition1}),~(\ref{StoppingCondition2}). 

\begin{table}[ht!!]
\footnotesize
\centering
\begin{tabular}  
{p{6mm} p{6mm} p{6mm} p{6mm} p{6mm} p{6mm} p{6mm} p{6mm} p{6mm} p{9mm} p{10mm}}
\hline
\diagbox[width=3em]{$\alpha$}{$\beta$} & 0 & 0.1 & 0.2 & 0.3 & 0.4 & 0.5 & 0.6 & 0.7 & 0.8 & 0.9 \\ 
\hline
0.05 & \cellcolor{orange!20} 2629 & 2365 & 2103 & 1845 & 1581 & 1319 & 1057 & 797 & 535 & 281 \\  
0.1  & 1315 & 1185 & 1053 & 923 & 793 & 661 & 533 & 401 & 273 & 89 (!) \\  
0.15 & 877 & 791 & 703 & 617 & 531 & 443 & 357 & 271 & 187 & 127 (!) \\  
0.2  & 659 & 595 & 529 & 463 & 399 & 333 & 269 & 205 & 145 & 99 (!) \\  
0.25 & 527 & 475 & 423 & 371 & 319 & 267 & 217 & 165 & 121 & 91 (!) \\  
0.3  & $-$ & $-$ & 355 & 311 & 267 & 225 & 181 & 141 & 107 & 141 (!) \\   
0.35 & $-$ & $-$ & $-$ & $-$ & 231 & 193 & 157 & 123 & 63 (!) & 123 (!) \\  
0.4  & $-$ & $-$ & $-$ & $-$ & $-$ & 187 & 141 & 109 & \cellcolor{green!20} 91 & 103 (!) \\ \hline
\end{tabular}   
\caption{\footnotesize For Case~3 of generating the Hadamard gate, the comparative results of GPM-1 with various fixed $\alpha$ (column with $\beta=0$) and of GPM-2 with various fixed $(\alpha, \beta)$. Each cell shows the numbers of the Cauchy problems which were solved for the corresponding $(\alpha,\beta)$ GPM case. For each cell with ``$-$'', the corresponding GPM cannot find an~acceptable result under the maximum number of iterations equal to~2000. The sign~``(!)'' means that the controls turn out to be jittery near $t=0,T$. \label{MorzhinPechenTable1}} 
\end{table}  
\normalsize 

\begin{figure}[ht!]
\centering
\includegraphics[width=\linewidth]{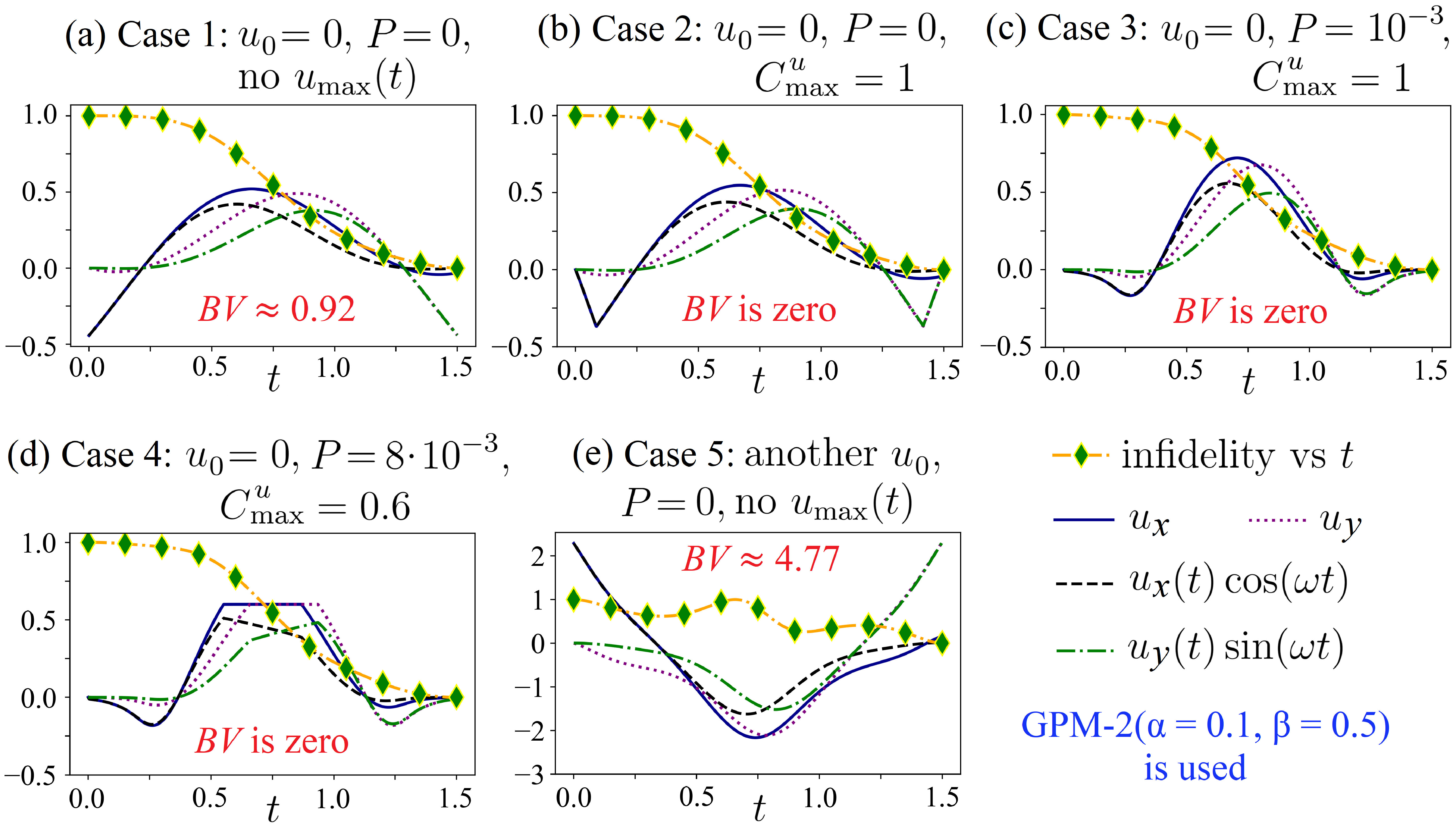}
\caption{\footnotesize For Cases 1--5 of generating the Hadamard gate, the infidelity vs~$t$, numerically optimized controls~$u_x, u_y$, combinations $u_x(t) \cos(\omega t)$, $u_y(t) \sin(\omega t)$, and $BV = \sum\nolimits_{l=1}^2 \left( |u_l(0)| + |u_l(T)| \right)$.}
\label{MorzhinPechenFig2} 
\end{figure} 

\begin{table}[ht!!]
\footnotesize
\centering
\begin{tabular}  
{p{57mm} p{12mm} p{12mm} p{16mm} p{17mm}}
\hline
\diagbox[width=17.5em]{GPM type}{Case} & 1 & 2 & 3 & 4 \\ \hline
\rowcolor{DefinedColor}
GPM-1($\alpha_k$, GSA) & 6,~1807 & 10,~3011 & 62,~18663 & 88,~26489 \\  
GPM-1($\alpha_k$, N.-M.) & 6,~237 & 4,~182 & 85,~3757 & 224,~7023 \\
\rowcolor{DefinedColor}
GPM-1($\alpha=0.1$) & 39,~79 & 53,~107 & 657,~1315 & 740,~1481 \\   
GPM-1($\alpha_k$, adapt., N.-M.),  $\theta=0.4$, with / without resetting to~$u^{\rm best}$ & 6,~192 / 6,~192 & 4,~181 / 4,~181 & 271,~756 / 275,~764 & 965,~2156 / 1166,~2558 \\
\rowcolor{DefinedColor}
GPM-2($\alpha=0.1,~\beta=0.5$) & 14,~29 & 21,~43 & 330,~661 & 379,~759 \\  
GPM-2($\alpha_k,~\beta_k$, GSA) & 7,~2108 & 6,~1807 & 17,~5118 & 15,~4516 \\  
\rowcolor{DefinedColor}
GPM-2($\alpha_k$, $\beta_k$, N.-M.) & 3,~266 & 3,~356 & 13,~1292 & 25,~2206 \\  \hline 
\end{tabular}  
\caption{\footnotesize For Cases 1--4 of generating the Hadamard gate, the comparative results of the various GPM forms. Each cell with numbers shows how many iterations (the numbers before ``,'') and solved Cauchy problems were spent.\label{MorzhinPechenTable2}}
\end{table} 
\normalsize 

\begin{figure}[ht!]
\centering
\includegraphics[width=\linewidth]{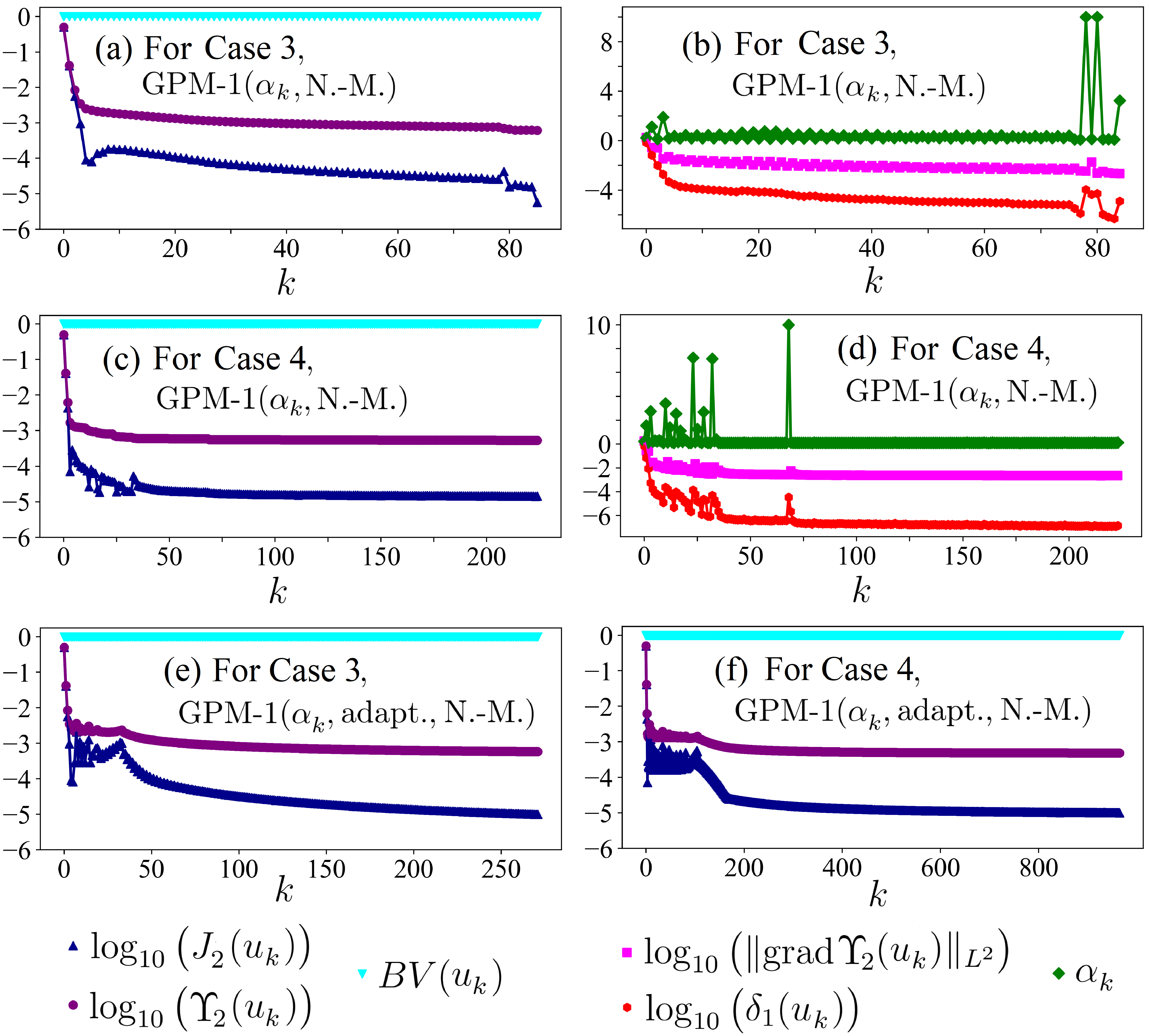}
\caption{\footnotesize For Cases~3,~4 of generating the Hadamard gate, the results of GPM-1 with adjusting $\alpha_k$ at each iteration and of the adaptive GPM-1.}
\label{MorzhinPechenFig3} 
\end{figure}

GPM-2($\alpha_k,\beta_k$, N.-M.) is taken with either the constraint $(\alpha,\beta) \in [0, 10] \times [0, 0.8]$ or the modified constraint $(\alpha,\beta) \in [0, 10] \times [0, 0.99 \, \overline{\beta}^{(k)}]$ with $\overline{\beta}^{(0)} = 0.9$. As~Table~\ref{MorzhinPechenTable1} shows, for the same Case~3, the largest and least complexities are~2629 and~91. In this table and also in Table~\ref{MorzhinPechenTable2}, we also see that, GPM-2 is able to {\it significantly accelerate} GPM-1 for the same~$\alpha$, and this effect is obtained only owing to the inertial term in GPM-2. Fig.~\ref{MorzhinPechenFig2} shows the efficient operation of GPM-2$(\alpha=0.1, \beta=0.5)$ and subfigures~(a,e) show the essentially different two numerically optimized vector controls corresponding to the different initial guesses. Moreover, Table~\ref{MorzhinPechenTable2} shows that it is possible to {\it more strongly decrease}~$\Upsilon_2$ at an~iteration by the cost of the problems of optimizing $\alpha$ or $(\alpha,\beta)$. For these problems, we do not aim to tune GSA or N.-M. method for a possible best reduction of the complexity of GPM-1(fixed $\alpha$), GPM-2(fixed $\alpha,\beta$) without decreasing their quality. The results of the adaptive GPM-1 are shown in Table~\ref{MorzhinPechenTable2} and Fig.~\ref{MorzhinPechenFig3}(e,f). We see that the resetting to $u^{\rm best}$ in the adaptive GPM-1 may be reasonable for reducing the GPM complexity. As Fig.~\ref{MorzhinPechenFig3}(a,c) show, $\Upsilon_2$ monotonically decreases when GPM-1($\alpha_k$, N.-M.) is used. Comparing subfigures~(a,e) for Case~3 and comparing subfigures~(c,f) for Case~4 show that the adaptive GPM-1 spent more iterations than GPM-1($\alpha_k$, N.-M.), but, as Table~\ref{MorzhinPechenTable2} shows, the adaptive GPM-1 spent less number of the solved Cauchy problems. 

In Table~\ref{MorzhinPechenTable1}, we see that GPM is robust in the sense that there are  60 pairs $(\alpha,\beta)$ shown in the table that the corresponding cells in Table~\ref{MorzhinPechenTable1} are without ``$-$'' and ``!'', i.e. GPM numerically solves the problem and the resulting controls are not jitter; the differences, as we note above, are in the GPM complexity, but it is important that any of these 60 GPM cases solves the problem. 

\subsection{Generating the CNOT gate for a~two-qubit superconducting system}
\label{subsection4.3}

Consider a two-qubit model representing superconducting qubit manipulated by a~quasiperiodic driving~\cite{RigettiPRL2005}. It corresponds to setting in~(\ref{InitProblemForSchrodingerEq}):
\begin{align*}
H_0 &= \Omega_1 {\rm Z}_1 + \Omega_2 {\rm Z}_2 + \widehat{\Omega} \, {\rm X}_1 {\rm X}_2, \\
\langle V(t), u(t) \rangle_{E^4} &= \sum\nolimits_{j=1} ^2 [u_j^x(t) \cos(\om_j t) + u_j^y(t) \sin(\om_j t)] {\rm X}_2,
\end{align*}
i.e. $V_1(t) = \cos(\omega_1 t) {\rm X}_2$, $V_2(t) = \sin(\omega_1 t) {\rm X}_2$, $V_3(t) = \cos(\omega_2 t) {\rm X}_2$, $V_4(t) = \sin(\omega_2 t) {\rm X}_2$ (quasiperiodic driving). Here $\widehat{\Omega} = |\Omega_1 - \Omega_2| / k$ (e.g., $k = 10$), the matrices $X_1 = X \otimes \mathbb{I}_2$, $X_2 = \mathbb{I}_2 \otimes X$, ${\rm Z}_1 = {\rm Z} \otimes \mathbb{I}_2$, and ${\rm Z}_2 = \mathbb{I}_2 \otimes {\rm Z}$ are the Pauli-X, Pauli-Z matrices for the 1st and 2nd qubits, control $u = (u_{x,1}, u_{y,1}, u_{x,2}, u_{y,2})$. Take $\Omega_1 = \omega_1 = 1$, $\Omega_2 = \omega_2 = 0.2$, $\widehat{\Omega} = 0.08$. Use $J_2(u)$ for the problem of generating the CNOT gate $W^{\rm CNOT} = 
\begin{pmatrix}  1 & 0 & 0 & 0 \\
0 & 1 & 0 & 0 \\
0 & 0 & 0 & 1 \\
0 & 0 & 1 & 0
\end{pmatrix}$. Take $T=20$ and consider the following two cases: 
\begin{itemize}
\item Case 1: $u^{(0)}=0$, $P=0$, without~(\ref{PointwiseConstraintForControl});

\vspace{-0.15cm}

\item Case 2: $u^{(0)}=0$, both (\ref{PointwiseConstraintForControl}) with $C_{\max}^u = 3$ and the integral term in $\Upsilon_2(u)$ are~used. 
\end{itemize} 

\begin{figure}[ht!]
\centering
\includegraphics[width=\linewidth]{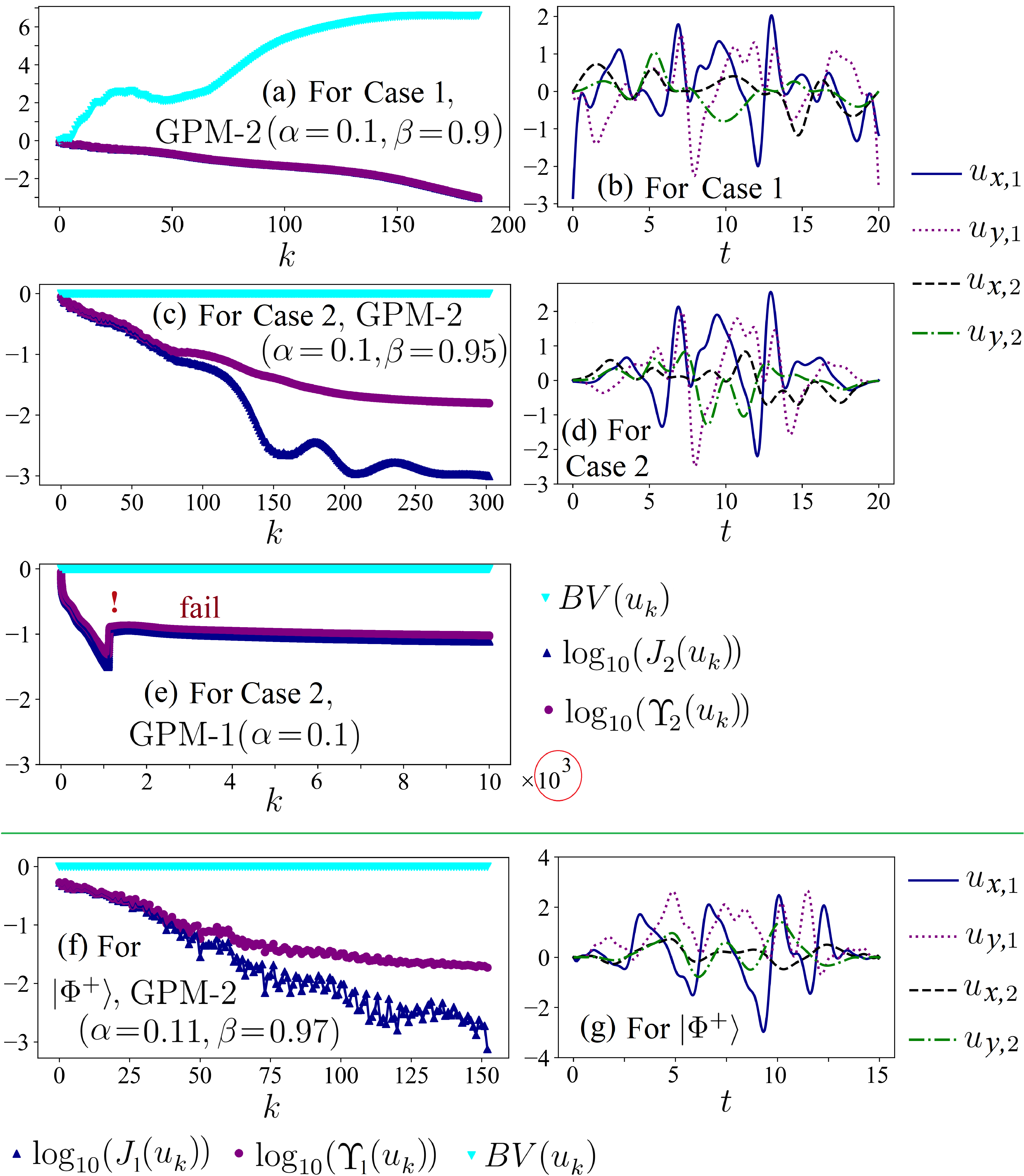}
\caption{\footnotesize Subfigures~(a)--(e): for Cases~1,~2 of generating CNOT. Subfigures~(f),~(g): for the problem of generating the Bell state $|\Phi^+\rangle$.} 
\label{MorzhinPechenFig4} 
\end{figure}

For both Cases~1,~2, Fig.~\ref{MorzhinPechenFig4} shows the results of GPM-2($\alpha=0.1, \beta=0.95$) and, in addition, of GPM-1($\alpha=0.1)$. For Case~1, where the requirement to switch-on at $t=0$ and switch-off at $t=T$ is not taken, subfigure~(b) shows that the numerically optimized~$u$ essentially violates this requirement. Subfigure~(d) shows that the resulting~$u$ satisfies this requirement used in Case~2. This aspect is also shown by $BV(u_k)$ vs~$k$ in subfigures~(a,c). We see the successful work of the GPM-2 with the fixed parameters. In subfigures~(a,e), we compare the $\log_{10}$ graphs for the objectives. Subfigure~(e) shows that, first, the successful part (with respect to decreasing values of $\Upsilon_2$) of the GPM-1 work spent essentially more iterations (and the solved Cauchy problems) than GPM-2 for the same effect under the same $\alpha=0.1$ and, second, the GPM-1 work fails near after one thousand iterations and, after that, cannot make up for failure even if we allow ten thousand iterations. Thus, the inertial term in GPM-2 is essential here. In Fig.~\ref{MorzhinPechenFig4}(b,d), we see that the numerically optimized $u_{x,1}, u_{y,1}$ have such the magnitudes that their absolute values are larger in comparison to such the values of the magnitudes of~$u_{x,2}, u_{y,2}$; thereby, we note that $\Omega_1 = \omega_1 = 1$ while $\Omega_1 = \omega_1 = 0.2$. 

\subsection{Generation of the Bell state \texorpdfstring{$| \Phi^+ \rangle$} for the two-qubit superconducting system}
\label{subsection4.4}

For the same closed two-qubit superconducting system, consider $C_{\max}^u = 5$ and the state transfer problem of the initial state $\psi_0 = |00\rangle$ to the target Bell state $\psi_{\rm target} = |\Phi^+\rangle = \frac{1}{\sqrt{2}} (|00\rangle + |11\rangle) = \frac{1}{\sqrt{2}} (1,0,0,1)^{\rm T}$, where $|00\rangle = (1,0,0,0)^{\rm T}$ and $|11\rangle = (0,0,0,1)^{\rm T}$. Use $\Upsilon_1(u)$ with $J_1(u)$ under (\ref{PointwiseConstraintForControl}). Take $T = 15$, $P = 10^{-4}$, $C_S =25$. The successful work of GPM-2($\alpha=0.11$, $\beta=0.97$) starting from $u_{0,l}(t) = 0.2 \,{\rm sinc}\Big(8 \pi \Big(\frac{t}{T} - \frac{1}{2} \Big)^3 \Big)$ is shown in Fig.~\ref{MorzhinPechenFig4}(f,g). 

\subsection{Generating the Werner state for a~two-qubit open system}
\label{subsection4.5}

Consider now operation of the GPMs for the following controlled open quantum system~\cite{MorzhinPechenEntropy2024}: 
\begin{align*}
H_0 &= \frac{\omega_1}{2} {\rm Z} \otimes \mathbb{I}_2 + \frac{\omega_2}{2} \mathbb{I}_2 \otimes {\rm Z}, \quad 
H_n = \Lambda_1 n_1(t) {\rm Z} \otimes \mathbb{I}_2 + \Lambda_2 n_2(t) \mathbb{I}_2 \otimes {\rm Z}, \\
V &= Q_1 \otimes \mathbb{I}_2 + \mathbb{I}_2 \otimes Q_2, \quad 
Q_j = \sin\theta_j \cos\varphi_j {\rm X} + \sin\theta_j \sin\varphi_j {\rm Y} + \cos\theta_j {\rm Z}, \\
\mathcal{D}_{n(t)}(\rho_t^c) &= \sum\nolimits_{j=1}^2 \Big[ \Omega_j \left( n_j(t) + 1 \right) \Big(2 \sigma_j^- \rho_t^c \sigma_j^+ - \left\{ \sigma_j^+ \sigma_j^-, \rho_t^c \right\}\Big) \nonumber \\
& \qquad + \Omega_j n_j(t) \Big(2 \sigma_j^+ \rho_t^c \sigma_j^- - \left\{ \sigma_j^- \sigma_j^+, \rho_t^c \right\} \Big) \Big],
\end{align*}
where $\varepsilon > 0$, the matrices $\sigma_1^{\pm} = \sigma^{\pm} \otimes \mathbb{I}_2$, $\sigma_2^{\pm} = \mathbb{I}_2 \otimes \sigma^{\pm}$ are obtained with $\sigma^+ = \begin{pmatrix}
	0 & 0 \\ 1 & 0
\end{pmatrix}$, 
$\sigma^- = \begin{pmatrix}
	0 & 1 \\ 0 & 0
\end{pmatrix}$; 
the parameters $\omega_j, \Lambda_j, \Omega_j$ and angles $\theta_j, \varphi_j$ are given. In this system, coherent control~$u$ is scalar and incoherent control is $n = (n_1, n_2)$. The two-qubit Werner state (e.g., in~\cite{RiedelGardingEtAlEntropy2021}) is defined for $p \in [0,1]$ as $\rho_W = 
\begin{pmatrix}
\frac{1-p}{4} & 0 & 0 & 0 \\
0 & \frac{1+p}{4} & -\frac{p}{2} & 0 \\
0 & -\frac{p}{2} & \frac{1+p}{4} & 0 \\
0 & 0 & 0 & \frac{1-p}{4}
\end{pmatrix}$, and is considered as target for the following two problems with $T=6$ and~(\ref{certainftc}): 
\begin{itemize}
\item keeping the system at $\rho_0 = \rho_W$ at the whole $[0, T]$ in the sense of~$\Theta_2(c)$;

\vspace{-0.15cm}

\item steering--keeping, i.e. steering from $\rho_0 \neq \rho_W$ to $\rho_W$ and keeping at the predefined rest part of $[0,T]$ in the sense of~$\Theta_1(c)$ with $\sigma = \{ 0, ~ t \in [0, \frac{1}{2}); ~ 2 t - 1, ~ t \in [\frac{1}{2}, 1]; ~ 1, ~ t \in (1, 6] \}$.
\end{itemize}

\begin{figure}[ht!]
\centering
\includegraphics[width=\linewidth]{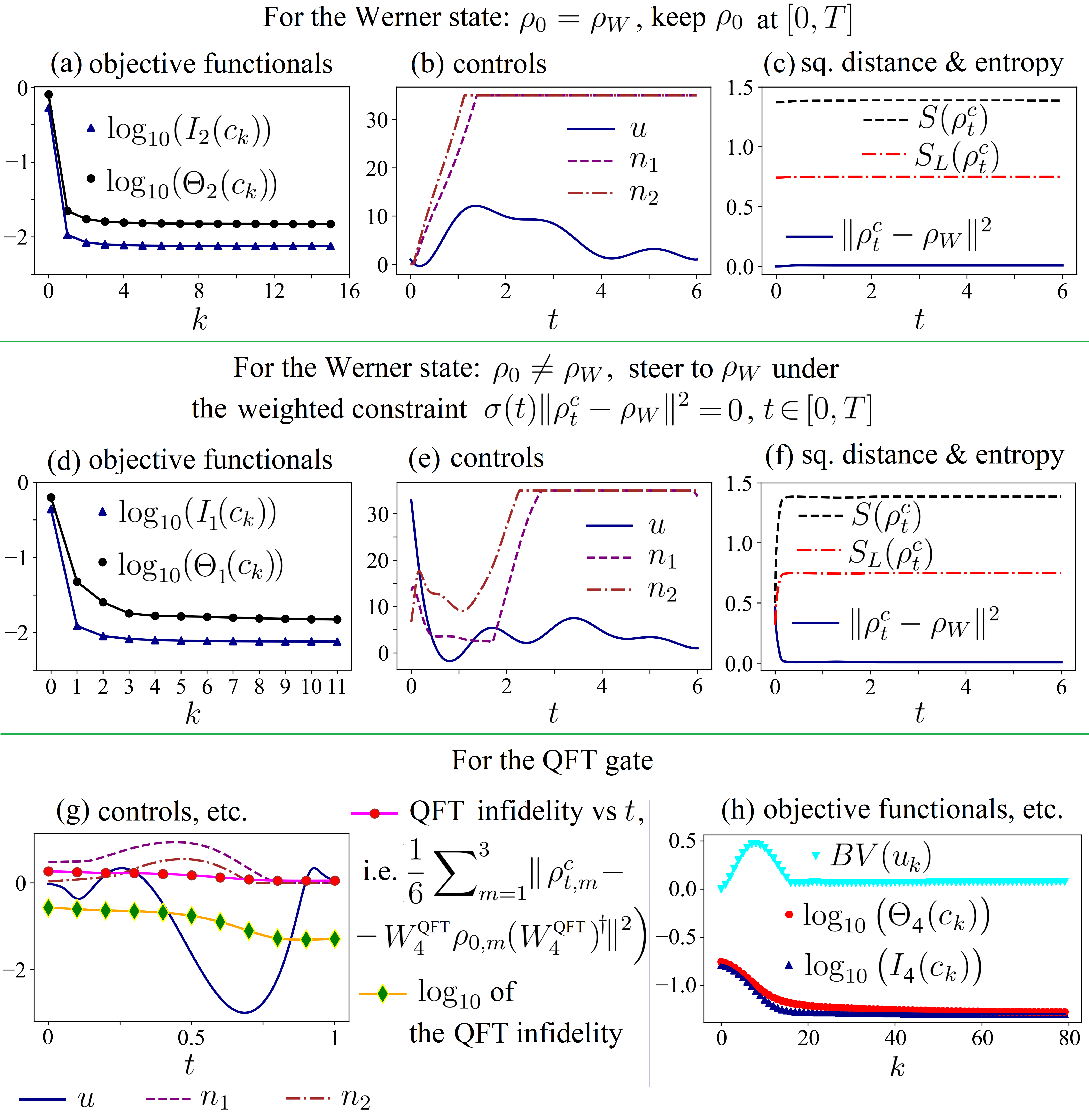}
\caption{\footnotesize For the open two-qubit system: (a--f)~GPM-2 results for the keeping (subfigures~(a,b,c)) and steering--keeping (subfigures~(d,e,f)) problems for the Werner state~$\rho_W$; (g,~h)~GPM-2 results for the QFT generation problem.}
\label{MorzhinPechenFig5} 
\end{figure}

Take $\varepsilon = 0.25$, $\omega_1 = 1$, $\omega_2 = 1.2$, $\theta_1 = \theta_2 = \frac{\pi}{4}$, $\varphi_1 = \varphi_2 = \frac{\pi}{2}$, $\Lambda_1 = 0.6$, $\Lambda_2 = 0.8$, $\Omega_1 = 0.8$, $\Omega_2 = 1$. In the constraint~(\ref{ParallelepipedalConstraintControlcOpen}), take $C_{\max}^u = 35$, $C_{\max}^n = 35$. Fix $p=0.1$ in $\rho_W$, so the von Neumann entropy $S(\rho_W) = - {\rm Tr}(\rho_W \log \rho_W) \approx 1.372$, i.e. near the maximal value $S_{\max} \approx 1.386$ for the case $N=4$. In the keeping problem, take $\rho_0 = \rho_W$ and $\Theta_2(c)$ with $P_{\rho}=1$, $P_u = P_n = 0$. In the steering--keeping problem, take $\rho_0 = {\rm diag}\left(0, 0, \frac{1}{5}, \frac{4}{5}\right)$, $\rho_{\rm target} = \rho_W$, and $\Theta_1(c)$ with $P_{\rho}=1$, $P_u = P_n = 0$.  GPM-2($\alpha=10$, $\beta=0.95$) starting from the initial guess $u_0 = 1$, $n_{1,0} =  n_{2,0} = 0$ reaches the stopping condition $I_j(c_k) \leq 0.008$~~\&~~$\Theta_j(c_k) \leq 0.015$ for $j \in \{1,2\}$ via 11 iterations (23 solved Cauchy problems) for the keeping problem and via 15 iterations for the steering--keeping problem. Fig.~\ref{MorzhinPechenFig5}(a--f) shows the successful results of such the GPM-2. Here linear entropy $S_L(\rho_t^c) = 1 - {\rm Tr}\big((\rho_t^c(t))^2\big)$ vs $t$ is also considered. 

\subsection{Generating the quantum Fourier transform gate for the two-qubit open system}
\label{subsection4.6}

For the same two-qubit open system, consider the problem of generating the quantum Fourier transform (QFT) gate $W_4^{\rm QFT} = \dfrac{1}{2} 
 \begin{pmatrix}
 1 & 1 & 1 & 1 \\
 1 & i & -1 & -i \\
1 & -1 & 1 & -1 \\
 1 & -i & -1 & i
\end{pmatrix}$. The article~\cite{PalaoKosloff2002} considers the problem of generating the QFT gate when the dynamics is unitary. Since our dynamics is dissipative, we use the Goerz--Reich--Koch (GRK) approach~\cite{GoerzReichKochNJP2014} adopted to our system with the objective~(\ref{GRKI4inf}) with $W_4^{\rm QFT}$. Consider $\Theta_4(c)$ with~(\ref{certainftc}). Fix $T=1$, $C_{\max}^u = 20$, $C_{\max}^n = 5$, $P_u = 10^{-4}$. GPM-2($\alpha=0.1, \beta=0.7$) starting from $u_0 = 2 \sin(2\pi t/T)$, $n_{1,0} = n_{2,0} = 0$ is used. Fig.~\ref{MorzhinPechenFig5}(g,h) show the obtained non-trivial controls, evolution of the infidelity $\frac{1}{6}\sum\nolimits_{m=1}^3 \| \rho^c_{t,m} - W_4^{\rm QFT} \rho_{0,m} (W_4^{\rm QFT})^{\dagger} \|^2$ vs~$t \in [0,T]$ and change of $I_4(c_k),~\Theta_4(c_k)$, $BV(u_k)$~vs~$k$. 

\subsection{Maximizing the Hilbert--Schmidt overlap for a qutrit open system}
\label{subsection4.7}

\begin{figure}[ht!]
\centering
\includegraphics[width=\linewidth]{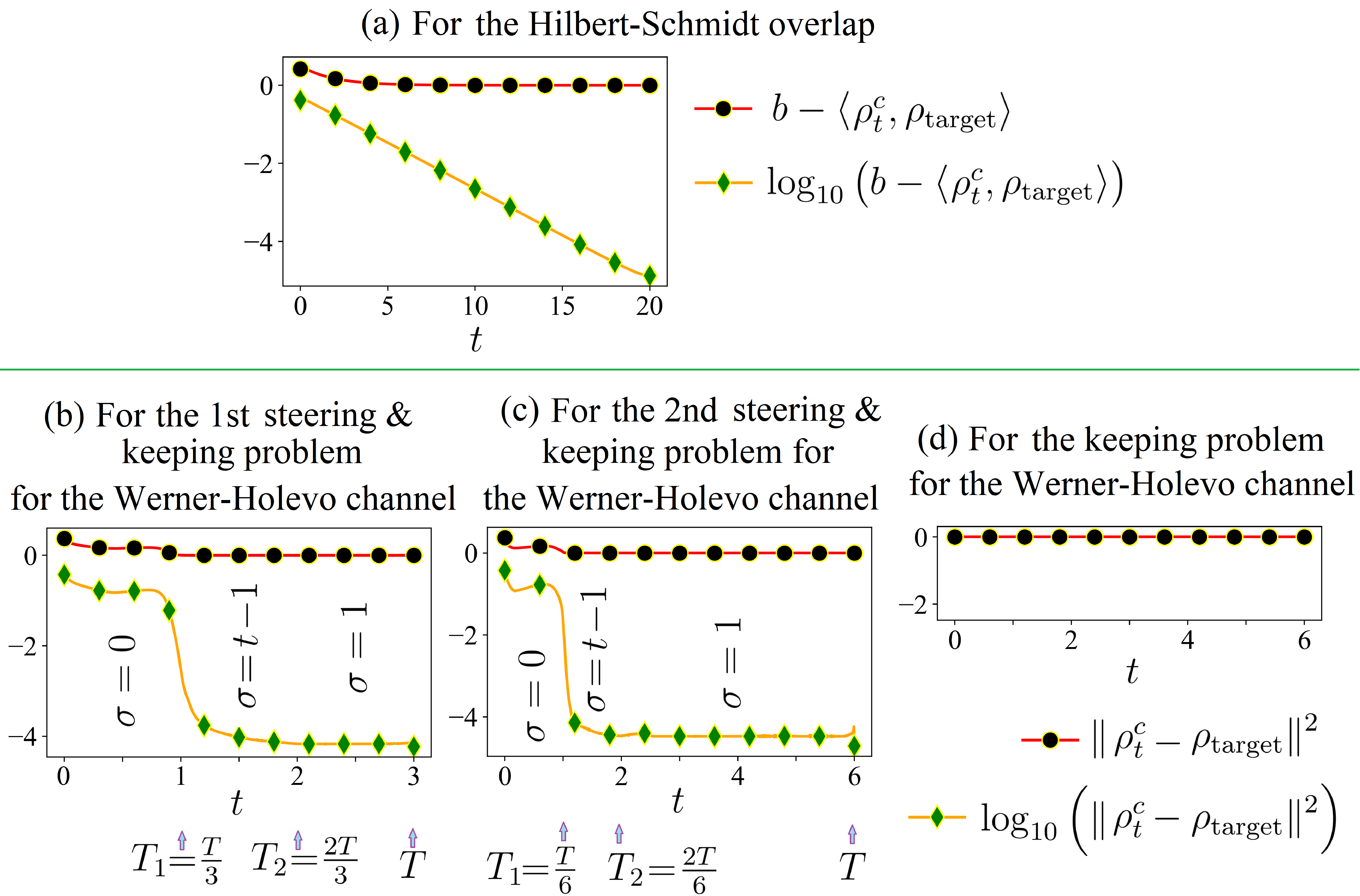}
\caption{\footnotesize GPM results for the open qutrit system: (a)~for maximizing the Hilbert--Schmidt overlap with~$\Theta_3(c)$; (b,c)~for the two steering--keeping problems for the Werner--Holevo channel (with the squared Hilbert--Schmidt distance); (d)~for the keeping problem for this channel (in terms of the distance).}
\label{MorzhinPechenFig6} 
\end{figure}

Following~\cite{MorzhinPechenProcSteklov2024}, consider the open $\Lambda$-system  with scalar control~$u$ and control $n = (n_1, n_2)$ ($N_n = 2$ for the two allowed directions in the $\Lambda$-system): 
\begin{eqnarray*}
& H_0 = \begin{pmatrix}
	0 & 0 & 0 \\
	0 & E_2 & 0 \\
        0 & 0 & E_3
\end{pmatrix}, \quad 
V = \begin{pmatrix}
	0 & 0 & V_{13} \\
	0 & 0 & V_{23} \\
        V_{13}^{\ast} & V_{23}^{\ast} & 0
\end{pmatrix}, 
\quad H_n \equiv 0, \quad \varepsilon = 1, \\
& \mathcal{D}_{n(t)}(\rho_t^c) = \sum\nolimits_{j=1}^2 \Big[ C_{j,3} \,(n_j(t)+1) 
\big( 2 A_{j,3} \, \rho_t^c \,A_{j,3}^{\dagger} - \{ A_{j,3}^{\dagger} \, A_{j,3}, \rho_t^c \} \big) \nonumber \\
& \qquad\qquad\qquad + C_{j,3} \, n_j(t) \big( 2 A_{j,3}^{\dagger} \, \rho_t^c A_{j,3} - \{ A_{j,3} \, A_{j,3}^{\dagger}, \rho_t^c \} \big)\Big],
\end{eqnarray*}
where $V_{13} = V_{13}^{\ast}$, $V_{23} = V_{23}^{\ast}$, the admissible transitions are determined by the matrices 
$A_{j,3} = V_{j,3} \, | j \rangle \langle 3|$, $j=1,2$, i.e.
$A_{13} = V_{13} \, | 1 \rangle \langle 3| = 
\begin{pmatrix}
0 & 0 & V_{13} \\ 
0 & 0 & 0 \\
0 & 0 & 0 
\end{pmatrix}$, $A_{13}^{\dagger} = 
\begin{pmatrix}
0 & 0 & 0 \\ 
0 & 0 & 0 \\
V_{13} & 0 & 0 
\end{pmatrix}$, 
$A_{23} = V_{23} \, | 2 \rangle \langle 3| = 
\begin{pmatrix}
0 & 0 & 0 \\ 
0 & 0 & V_{23} \\
0 & 0 & 0 
\end{pmatrix}$, 
$A_{23}^{\dagger} = \begin{pmatrix}
0 & 0 & 0 \\ 
0 & 0 & 0 \\
0 & V_{23} & 0 
\end{pmatrix}$. Take $E_2 = 1$, $E_3 = 2.5$, $V_{13} = 1$, $V_{23} = 1.7$, $C_{13}=0.4$, $C_{23}=0.2$, $\rho_0 = {\rm diag}\left(\frac{7}{10}, \frac{3}{10}, 0 \right)$, $\rho_{\rm target} = {\rm diag}\left(\frac{1}{5}, \frac{4}{5}, 0 \right)$. Use $\Theta_3(c)$ with~(\ref{certainftc}), $\sigma = \{ 0, ~ t \in [0, 0.7 \,T); ~ 1, ~ t \in [0.7 \,T, T]\}$, $T = 20$, the upper bound $b = 0.8$, $C_{\max}^u = 20$, $C_{\max}^n = 20$,  $P_{\rho}=5$, $P_u = 10^{-5}$, $P_n = 10^{-3}$. Using GPM-2(fixed $\alpha, \beta$) shows that the integral state term in $\Theta_3(c)$ allows to help to GPM in  better achieving the upper bound by the final overlap. Fig.~\ref{MorzhinPechenFig6}(a) shows how the infidelity changes vs~$t$ at the numerically optimized controls in the case when the integral state term is used. Moreover, the problem with the state constraint is analyzed with various final times at a~decreasing sequence $\{T_j\}$. The computations show that the resulting values of the final overlap depend on~$T$ (time is as a~control resource) and that $T=20$ provides $0.8 - \langle \rho_T^c, \rho_{\rm target} \rm \rangle \approx 10^{-5}$ and the von Neumann entropy $S(\rho_T^c) \approx 3 \cdot 10^{-4}$.

\subsection{Steering--keeping and keeping problems for a qutrit Werner--Holevo channel}
\label{subsection4.8}

The Werner--Holevo channel~\cite{WernerHolevoJMathPhys2002,RoofehKarimipourPRA2024} is important for the theory of quantum information and computing and is defined by 
${\rm WHC}(\sigma) = \frac{1}{N-1} \left({\rm Tr}(\sigma) \mathbb{I}_N - \sigma^{\rm T} \right)$ for  a~matrix~$\sigma$ whose order is~$N>1$ and trace is~1 (but not necessarily a density matrix). For $N = 3$, consider the given above qutrit system and the following three problems under the constraint~(\ref{ParallelepipedalConstraintControlcOpen}) and with~(\ref{certainftc}):
\begin{itemize}
\item 1st steering--keeping problem where $\rho_0 = {\rm diag}\left(0, \frac{1}{2}, \frac{1}{2} \right)$, $\rho_{\rm target} = {\rm WHC}(\rho_0) = {\rm diag}\left(\frac{1}{2}, \frac{1}{4}, \frac{1}{4} \right)$, $T=3$; the goal is to steer to $\rho_{\rm target}$ and further keep at this state in the sense of~$\Theta_1(c)$ with $\sigma = \{ 0, ~ t \in [0, 1); ~ t - 1, ~ t \in [1, 2]; ~ 1, ~ t \in (2, 3] \}$; 

\vspace{-0.15cm}

\item 2nd steering--keeping problem which differs from the previous problem so that $T=6$ and the time range for the system keeping is significantly longer in the sense of $\sigma = \{ 0, ~ t \in [0, 1); ~ t - 1, ~ t \in [1, 2]; ~ 1, ~ t \in (2, 6] \}$; 

\vspace{-0.15cm}

\item keeping problem where $T=6$ and the goal is to keep at $\rho_0 = {\rm diag}\left(\frac{1}{2}, \frac{1}{4}, \frac{1}{4}\right)$, i.e. at the given above ${\rm WHC}({\rm diag}\left(0, \frac{1}{2}, \frac{1}{2})\right)$, at $[0,T]$ in the sense of~$\Theta_2$. 
\end{itemize}

For the both steering--keeping problems, take  $C_{\max}^u = 30$, $C_{\max}^n = 30$. For the  1st steering--keeping problem, use $\Theta_2(c)$ with $P_{\rho} = 30$ and some small $P_u, P_n >0$. For this problem, Fig.~\ref{MorzhinPechenFig6}(b) shows how the obtained by GPM-3(fixed $\alpha, \beta, \gamma$) dependence $\| \rho_t^c - \rho_{\rm target}\|^2$ of~$t$ at $[0,T]$ handles with the weighting  function~$\sigma$. Further, the 2nd steering--keeping problem is considered with $P_{\rho} = 20$. For this problem, Fig.~\ref{MorzhinPechenFig6}(c) shows the corresponding result. For the keeping problem, take $P_{\rho} = 10$, the corresponding result of GPM-3(fixed $\alpha, \beta, \gamma$) is shown in~Fig.~\ref{MorzhinPechenFig6}(d). In these examples, GPM gives the expected~results.

\section{Conclusions}
\label{sectionConclusions}

In this work we provide an adaptation of the Gradient Projection Method (GPM) to various optimization problems for general $N$-level closed and open quantum systems, latter in general with simultaneous coherent and incoherent controls, with constraints imposed on the controls. The main advantage of the GPM  approach is its ability to {\it exactly} satisfy the constraints imposed on the controls. For these general $N$-level optimal control problems, we derive the adjoint systems and gradients of the objective functionals, the projection forms of the linearized versions of the Pontryagin maximum principle (Statements~\ref{statement1}--\ref{statement4}), and the various adapted  forms of GPM (one- and two-step, etc.) For comparison of the efficiency of various forms of the GPM, we consider several control problems with objective functionals to be minimized and with constraints on the controls, including the possible requirement for coherent control to continuously switch-on at $t=0$ and switch-off at $t=T$. As examples we consider in the numerical experiments generation of target states or unitary gates (e.g., H, CNOT, QFT, qutrit Werner--Holevo channel) for one- and two-qubit superconducting systems and two-qubit and qutrit open systems. 

The described above numerical experiments show that GPM may be a~useful optimization tool for a constrained quantum control optimization. For handling the initial- and end-point constraints on coherent control, we find  that the proposed approach, which combines the ordinary approach with adding weighted integral and the idea to use the corresponding time-dependent bounds for controls, is more efficient for optimization than the ordinary approach. 
In addition to the qualitative comparison of the five GPM forms in subsection~\ref{subsection2.2}, the numerical experiments indicate that GPM-1(fixed~$\alpha$) and GPM-2(fixed~$\alpha,\beta$) may be efficient with  a suitable tuning of their parameters and that GPM-2(fixed~$\alpha,\beta$) may be essentially better (faster and more reliable) than GPM-1(fixed~$\alpha$) with the same $\alpha$, as~Tables~\ref{MorzhinPechenTable1},~\ref{MorzhinPechenTable2}, and Fig.~\ref{MorzhinPechenFig4}(c,e)~show. Because Table~\ref{MorzhinPechenTable1} shows such the pairs of $\alpha,\beta$ that the corresponding values of the  complexity of GPM-2(fixed~$\alpha,\beta$) essentially differ from each other (in~26 times) and, moreover, the resulting control profiles may depend on these parameters, we conclude on the importance of trying various pairs of $\alpha,\beta$. 

\section*{Funding} 

For the control problems with unitary dynamics (section~2, subsections~4.2--4.4), the work was carried out with financial support from the University of Science and Technology MISIS, grant No.~K2-2022-025 in the framework of the Federal Academic Leadership Program ``Priority 2030''. For the control problems with non-unitary dynamics (section~3, subsections~4.5--4.8), the work was performed at the Steklov Mathematical Institute under the Russian Science Foundation grant No.~22-11-00330 (https://rscf.ru/en/project/22-11-00330/) except the work for the problem~(\ref{GRKI4inf}) (including subsection~4.6).

\end{document}